# Heat transport in crystalline organic semiconductors: coexistence of phonon propagation and tunneling

Lukas Legenstein[1], Lukas Reicht[1], Sandro Wieser[1,2], Michele Simoncelli[3,4], and Egbert Zojer[1, †]

[1] Institute of Solid State Physics, Graz University of Technology, NAWI Graz, Petersgasse 16, 8010 Graz, Austria
[2] Institute of Materials Chemistry, TU Wien, Getreidemarkt 9, 1060 Wien, Austria
[3] Theory of Condensed Matter Group, Cavendish Laboratory, University of Cambridge (UK)
[4] Department of Applied Physics and Applied Mathematics, Columbia University, New York (USA)

† corresponding author: egbert.zojer@tugraz.at

Understanding heat transport in organic semiconductors is of fundamental and practical relevance. Therefore, we study the lattice thermal conductivities of a series of (oligo)acenes, where an increasing number of rings per molecule leads to a systematic increase of the crystals' complexity. Temperature-dependent thermal conductivity experiments in these systems disagree with predictions based on the traditional Peierls-Boltzmann framework, which describes heat transport in terms of particle-like phonon propagation. We demonstrate that accounting for additional phonon-tunneling conduction mechanisms through the Wigner Transport Equation resolves this disagreement and quantitatively rationalizes experiments. The pronounced increase of tunneling transport with temperature explains several unusual experimental observations, such as a weak temperature dependence in naphthalene's conductivity and an essentially temperature-invariant conductivity in pentacene. While the anisotropic conductivities within the acene planes are essentially material-independent, the tunneling contributions (and hence the total conductivities) significantly increase with molecular length in the molecular backbone direction, which for pentacene results in a surprising minimum of the thermal conductivity at 300K.

## Introduction

Organic semiconductors (OSC) are applied in a variety of devices[1], like thin-film transistors[2], light-emitting diodes[3], or organic solar cells[4]. Currently they dominate the market for high-end displays in mobile phones, tablets, and TVs.[5] Therefore, most research on these materials has focused on their optical, electronic, and charge transport properties. In contrast, thermal properties of organic semiconductors and, here, especially heat transport has been studied to a much lesser degree.[6] Thus, there is a profound lack of detailed understanding of the underlying mechanisms at an atomistic level. In pristine and crystalline materials, heat transport is dominated by phonons, which can be rationalized by the Wiedemann Franz law, as electrical conductivities of undoped organic semiconductors are extremely small. Structurally, a peculiarity of OSCs is that one is not dealing with an extended 3D network of covalent bonds. Rather, covalently bonded molecules are held together by weak non-covalent van der Waals forces and electrostatic multipole interactions. The weak intermolecular interactions distinctly distinguish OSCs (and molecular crystals in general) from covalently-bonded semiconductors like Si or Ge, to which we will refer to as "conventional crystals". For the latter, thermal conduction has been studied much more thoroughly, both experimentally and by means of computer simulations.[7–10]

Still, computational modelling approaches analogous to those used for describing heat transport in conventional crystals have also been applied to OSCs. Here, the focus has been on molecular dynamics (MD)-based techniques, including approach-to-equilibrium MD (AEMD),[11–13] non-equilibrium MD (NEMD),[14,15] and Green-Kubo type equilibrium MD.[16] In recent years, these studies primarily dealt with showing the effect of defects or of structural modifications to the molecular backbones. Typically, they provide only phenomenological explanations for the underlying heat transport mechanisms. The main findings of the MD-based simulations and of the (accompanying) experimental works are that (i) the room temperature thermal conductivity $\kappa$ in OSCs is typically ultra-low ($\kappa <$ 1 Wm$^{-1}$K$^{-1}$)[11–15] and (ii) the thermal conductivity can be tuned by modifications to the molecular backbones, with the structural complexity and degrees of freedom of the substituents also playing an important role.[11,12,15]

MD-based methods predict heat transport by describing macroscopic heat currents and temperature gradients based on the average atomic motions in real space. This usually involves hundreds of thousands or even millions of time steps for which the equations of motions have to be solved for supercells containing (tens) of thousands of atoms. A conceptionally different approach, which to the best of our knowledge has not been applied to OSCs, is anharmonic lattice dynamics (ALD). In ALD, heat transport is described via the propagation of the energy quanta of excited lattice vibrations in reciprocal space. These are referred to as phonons. A common approach is to apply the (Peierls-)Boltzmann transport equation (BTE)[17], where phonons are semi-classically treated as propagating particles that transport heat. This is particularly useful for analyzing molecular crystals, as individual vibrational modes contributing to the transport can be classified based on their inter- or intramolecular character.[18,19]

The occupation of phonon modes obeys the Bose-Einstein statistics, which can be straightforwardly accounted for in ALD. Thus, this approach is better suited for describing heat transport at low temperatures compared to MD-methods, which rely on classical statistics (i.e., the equipartition theorem).[20] This is relevant also in the present context, where the temperature dependence of the thermal conductivity will be an important aspect. In view of what has been said above, it is not surprising that ALD has produced accurate predictions of thermal conductivities over a wide temperature range for many conventional crystals[21–23]. In these systems, phonons are well-defined states with small (spectral) linewidths compared to the occurring phonon band gaps.

However, the BTE systematically underestimates heat transport[24–26] in systems in which low, often temperature-invariant ("glass-like") thermal conductivities arise from strong anharmonicities and/or complex crystal structures. Notably, a similar situation prevails in OSC crystals, where broad phonon linewidths (due to significant anharmonicities) and narrow phonon band gaps (due to large unit cells and/or disorder) occur. These can lead to a significant overlap of broadened phonon bands, especially at increased temperatures. In such a case, the picture of particle-like phonon transport alone is insufficient and needs to be augmented. This can be done using the Wigner Transport Equation (WTE)[27], a generalization of the BTE, formulated as a unifying theory for thermal transport in both crystalline and amorphous solids.[28] While the BTE describes only the particle-like propagation of phonon wave packages, the WTE further includes transport arising from inter-mode coupling due to overlapping phonon bands. In the past, this so-called "wave-like" tunneling mechanism has been shown to provide crucial contributions to the thermal conductivities of highly anharmonic materials, such as perovskites[28–32] and skutterudites[26]. It is also important in materials with increased structural complexity and low anharmonicity, such as the hybrid polymorphs of silica[33] or alloys with compositional disorder used, e.g., for thermal barrier coatings.[34]

Investigating the effects of phonon tunneling in OSCs, a class of materials that is important for both technological applications and scientific research, will be at the heart of the present study. Additionally, we will also explore the impact of varying the structural complexity of the studied materials. While in previous works the complexity of materials has been increased by introducing compositional disorder[34,35], here, we will instead systematically increase the size of the molecular building blocks in the series of (oligo)acenes. This avoids breaking the long-range order in the studied molecular crystals, but at the same time significantly increases the number of atoms in the unit cell and, therefore, drastically increases the number of vibrational modes. A detailed description of the crystallographic structure of the acenes will be provided in the Results section. In contrast to the low anharmonicity in the polymorphs of silica with varying degrees of disorder,[33] we here focus on a class of highly anharmonic materials with room-temperature thermal conductivities $\kappa \ll 1$ Wm$^{-1}$K$^{-1}$. For the present study, it is particularly useful that a systematic discussion of structure-to-property relations for the low-frequency vibrations in acenes and of the associated phonon band structures has already been provided in ref. 18. There it has also been explicitly shown that the density of phonon bands in the low frequency region increases sharply with the number of rings in the acenes. In the current work we build on these insights, and extend them by going beyond the harmonic approximation, showing how increasing the molecular length (while preserving the molecular packing motif) influences thermal transport. In particular, we will dissect the thermal conductivities into their propagation and tunneling contributions, analyzing when one has to go beyond mere particle-like propagation and include wave-like tunneling mechanisms.

In doing that, we find for the series of acene crystals: (i) even at temperatures down to 100 K, the thermal conductivities in all acenes hardly exceed 1 Wm$^{-1}$K$^{-1}$; (ii) while thermal conductivities are virtually independent of the molecular lengths within the herringbone planes, they distinctly increase parallel to the molecular axes with molecular length; (iii) the quantitative agreement between our predictions and experiments for the temperature-dependent thermal conductivity of naphthalene and pentacene is excellent, but only when also considering wave-like phonon tunneling processes; (iv) this includes the observation that for most of the considered temperature range, the thermal conductivity displays a weaker temperature dependence than the classical $T^{-1}$ trend.[36,37] Besides the materials' high Debye temperatures, this is a consequence of wave-like phonon tunneling significantly increasing with temperature, which for tetracene and pentacene even results in a predicted increase of the thermal conductivity at temperatures exceeding ~350 K.

## Results

### Crystal structures of acenes

As mentioned above, the systems studied here are the aromatic acenes, which are π-conjugated molecules consisting of (linearly) fused benzene rings forming crystalline structures. The two- to five-ring systems are referred to as naphthalene[38], anthracene[39], tetracene[40], and pentacene[41]. Especially pentacene (and its derivatives) have traditionally been the prototypical molecular materials for realizing organic electronic devices.[42–44]

All four considered acenes crystallize in a herringbone structure with two molecules in the unit cell. This is shown in Fig. 1 for naphthalene (a) and pentacene (b). The herringbone arrangement of the molecules in a plane parallel to the (001) plane of the crystal (spanned by the $a_1$ and $a_2$ vectors) arises from the quadrupole interactions between the two molecules. The long axes of the molecules are roughly aligned with the lattice vector $a_3$, whose length increases with the number of rings, as shown in Fig. 1 (c) and (d). In the two shorter acenes, $a_1$ and $a_2$ are perpendicular to each other such that the molecular crystals are monoclinic, while tetracene and pentacene display a triclinic structure, with the angle between $a_1$ and $a_2$ amounting to roughly 86°.

In passing we note that for organic crystals typically a variety of polymorphs are identified; thus, the exact crystal structures[38–41] studied in this work need to be unambiguously specified, which is done in Supplementary Table 1. Unless otherwise stated, the unit cells in all following simulations were fixed to the DFT calculated cells (Supplementary Table 2). As a consequence, the impact of thermal expansion of the acene crystals on the temperature-dependent thermal conductivities will not be reflected in the majority of the presented data, as it is not in the focus of the present study. Still, it is briefly discussed in Supplementary Section 4.2. There we explicitly show that the trend for the temperature dependence of naphthalene's lattice thermal conductivity is not impacted by choosing the experimental unit cell obtained at 295 K instead of the DFT-relaxed cell. Moreover, its absolute value at room temperature is only 8% lower than when using the DFT-relaxed cell. As the deviations increase for low temperatures, picking the DFT-relaxed (i.e., the low-temperature) cell for the entire considered temperature range is more appropriate.

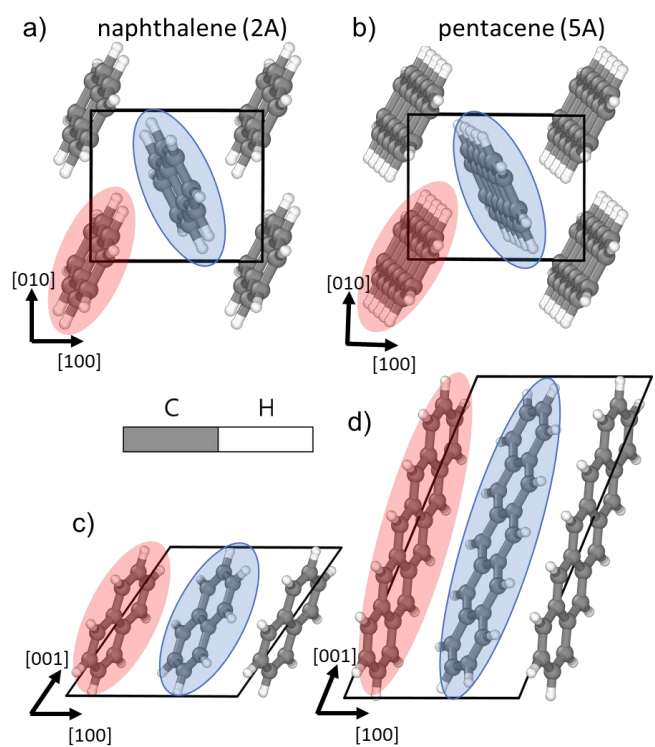

*Fig. 1: Crystal structures of naphthalene (2A) and pentacene (5A). To distinguish the two inequivalent molecules per unit cell, they are marked by red and blue ellipses. The structures are viewed along the [001] axes in (a) for 2A and (b) for 5A, showing the herringbone stacking and along the [010] axes in (c) and (d). The structures are visualized with Ovito.[45] Similar illustrations of anthracene and tetracene can be seen in Supplementary Fig. 1.*

### Moment tensor potentials as an efficient surrogate model for DFT

Usually, the phonon properties used to solve the WTE (or BTE) are calculated with forces from ab initio methods like density functional theory (DFT). However, when applying converged settings for basis sets, k-point grids, and supercells, especially the calculation of higher-order force constants becomes computationally intractable for the materials considered here (see Supplementary Section 2.3.). This calls for the use of computationally more efficient methods, such as density-functional tight binding (DFTB) and classical force fields (e.g. GAFF and COMPASS). These methods, however, already lack the necessary accuracy in describing harmonic phonon properties, as explicitly shown in ref. 46 for naphthalene.

Here, we overcome these limitations regarding computational costs and accuracy by employing a recently described machine-learning based workflow[47] to parametrize system-specific potentials on ab-initio force, energy, and stress data. This provides us with a suitable surrogate model for DFT. In particular, we use computationally extremely efficient Moment Tensor Potentials (MTP)[48] to calculate the atomic forces required to simulate thermal conductivities. These potentials have already been applied successfully for that purpose in previous studies.[49–54] Here, the MTPs are trained on relatively small sets of highly accurate dispersion-corrected DFT calculations, obtained during active-learning MD runs. The conditions for these MD simulations have been chosen to improve the accuracy for describing the comparatively small atomic displacements encountered when calculating second- and third-order force constants.[55] Further details on the MTP parametrization can be found in the Methods section. As shown in previous studies on metal-organic frameworks and polymeric materials[47,55], the obtained MTPs yield phonon properties at essentially DFT accuracy at computational costs reduced by at least four orders of magnitude.

## Validating the parametrized MTPs for describing phonon properties

The essential components to solve the BTE (and WTE) are phonon frequencies, group velocities, heat capacities, and lifetimes.[7,17,56,57] To calculate these phonon properties, harmonic and anharmonic interatomic force constants are required. As explained above, this is achieved for the structurally rather complex OSC crystals considered here by applying machine-learned MTPs. For the results to be meaningful, the accuracy of the force calculations must be very high. Therefore, as a first step the accuracy of the MTPs needs to be assessed by calculating phonon properties that are also accessible to DFT.

The MTPs were trained separately for each acene to maximize their accuracy. To validate each of these potentials, we compared phonon frequencies from MTPs with reference calculations performed with DFT. For these reference calculations taken from ref. 18, a

DFT methodology equivalent to the one used for training the MTPs was applied. The DFT relaxed crystal structures and the second-order force constants for calculating the phonon band structures were obtained from the NOMAD repository.[58]

As the fitting of the MTP parameters is a stochastic process, for each system, three MTPs were independently parametrized and the one with the smallest deviations between the MTP- and DFT-calculated phonon band structures was retained. In this context, it is worth mentioning that the frequency root-mean-square deviations (RMSDs) between the MTP and DFT calculated band structures are very small even for the nominally worst MTPs, as shown in Supplementary Table 5.

Fig. 2 illustrates that the MTPs for naphthalene (a) and pentacene (b) yield phonon band structures and densities of states that are practically indistinguishable from the respective DFT counterparts. For the sake of clarity, the phonon bands are shown in the energy range up to 20 THz that is most relevant for heat transport. Analogous plots for anthracene and tetracene can be seen in Supplementary Fig. 4 and Supplementary Fig. 5.

At a more quantitative level, the frequency RMSDs in Supplementary Table 5 testify to the exceptional performance of the MTPs with values as low as 0.073 THz (2.4 cm$^{-1}$) for the best MTPs. The deviations are particularly small in the low-frequency range defined as containing modes up to 5 THz, where they range between 0.034 THz (1.1 cm$^{-1}$) and 0.054 THz (1.8 cm$^{-1}$). These modes mostly correspond to intermolecular vibrations, which (due to their rather significant thermal occupation) are particularly relevant for heat transport. To further validate the accuracy of MTP against DFT, in Supplementary Fig. 6 we show that also the DFT- and MTP- calculated vibrational eigenvectors at the $\Gamma$-point are overall in good agreement.

A DFT calculation of higher-order force constants using converged numerical settings is beyond the rather significant computational capacities at our disposal (see Supplementary Section 2.3). This prevents a direct comparison of the acenes' phonon lifetimes between DFT and MTP. Instead, we tested the MTPs on a validation set[59] of DFT-calculated forces obtained from independent active-learning runs (detail in Supplementary Section 2.2.1). As the atomic displacements encountered in these runs are not restricted to the harmonic region, these forces also (implicitly) sample higher-order force constants. The correlation between DFT and MTP calculated forces is close to perfect, as can be seen in Supplementary Fig. 3.

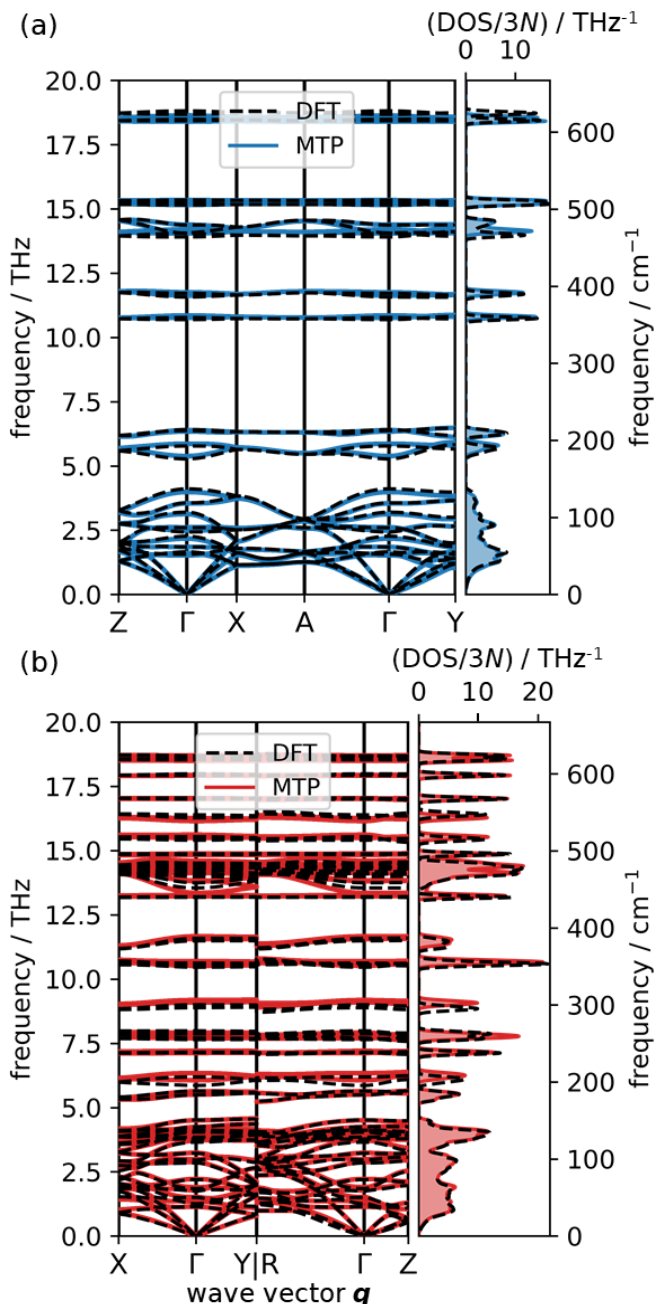

(a)

(b)

*Fig. 2: Phonon band structure and density of states of naphthalene (a) and pentacene (b) calculated with PBE+D3BJ (black, dashed line; data adopted from ref. 18) and with a system-specific MTP (colored, solid lines). Notably, the lengths of the high symmetry paths are not identical for different materials. Data from ref. 18 have been published by the Royal Society of Chemistry; copyright, the authors.*

Notably, the corresponding force RMSDs (listed in Table 1) are below 10 meVÅ$^{-1}$ for each material. These values are quite close to the 5 meVÅ$^{-1}$ threshold found by Póta et al.[60] to be sufficient for achieving an agreement within 2% between thermal conductivities calculated with DFT and a machine-learned interatomic potentials for the case of binary inorganic compounds. Due to the particular sophistication of the MTPs used here (see MTP "level" discussed in the

Method section), our force RMSDs are also well below what has been reported in several studies where MTPs successfully reproduced thermal conductivities from DFT. These studies focus on various classes of fundamentally different materials, which include semiconductors[50–52], 2D materials[51–53], CoSb$_3$ skutterudite[49], and different polymorphs of Ga$_2$O$_3$[54]. In passing we note that for certain materials also MTP-calculated 4$^{th}$-order force constants have been taken into consideration.[61,62] With regards to (ultra-)thermal conductivities, MTPs have already been employed to solve the WTE for materials like Zintl-phase SrCuSb and Sr$_2$ZnSb$_2$[63], La$_2$Zr$_2$O$_7$[64], and Cu$_7$PS$_6$ argyodite[65] just to name a few. Moreover, the workflow to parametrize the MTPs in this work is based on an earlier study of heat transport in the metal-organic framework material MOF-5[47] employing non-equilibrium molecular dynamics simulations. In that study, the MTPs excellently reproduced available experimental data.

*Table 1: Force validation errors for three independent MTPs for each acene. Shown are the total root-mean-square deviations between DFT and MTP forces for 200 structures from a validation set, obtained through active learning and independent of the training data used to parametrize the MTPs (see Methods section). The MTPs highlighted in bold yielded the lowest errors as well as the best phonon band structures and, thus, were retained throughout the rest of the study.*

| | force RMSD / meVÅ$^{-1}$ | | |
|---|---|---|---|
| acene | MTP #1 | MTP #2 | MTP #3 |
| 2A | 5.67 | **4.93** | 6.60 |
| 3A | 12.31 | 9.23 | **8.39** |
| 4A | 11.68 | **8.83** | 12.1 |
| 5A | **7.69** | 7.73 | 7.96 |

## Wigner formulation of the thermal conductivity

The Wigner formulation of thermal transport allows to describe not only the particle-like propagation of phonons, but also their wave-like tunneling arising from inter-mode coupling. In that sense, the more commonly used BTE can be regarded as a special case of the WTE with the BTE neglecting tunneling transport.[27] In passing we note that for describing the anharmonicities only third-order force constants are considered. This approximation is employed because rigorously accounting for anharmonicities beyond the third order in the context of the WTE is an open, non-trivial problem.[27,66] In practice, the accuracy of such an approximation is estimated in Supplementary Fig. 24 by rescaling the phonon linewidths to either increase or decrease the anharmonicity. There, we show that, due to the presence of dense phonon bands and significant third-order anharmonicities, the conductivity is only weakly affected by approximations in the perturbative treatment of the anharmonicity.

In the WTE, the total phonon thermal conductivity is the sum of two contributions: the particle-like transport from phonon propagation, which gives rise to the "populations" conductivity ($\kappa_P$), and the wave-like tunneling contribution from "coherences" ($\kappa_C$), which becomes more significant as anharmonicity or disorder increases.[27,28] Overall, this yields the following equation for the thermal conductivity tensor describing the relation between the heat flow in cartesian direction $\alpha$ driven by a temperature gradient in direction $\beta$.

$$\kappa_{tot}^{\alpha\beta} = \kappa_P^{\alpha\beta} + \frac{1}{VN_c} \sum_{q,s \neq s'} \frac{\omega(\boldsymbol{q})_s + \omega(\boldsymbol{q})_{s'}}{4} \left[ \frac{C(\boldsymbol{q})_s}{\omega(\boldsymbol{q})_s} + \frac{C(\boldsymbol{q})_{s'}}{\omega(\boldsymbol{q})_{s'}} \right]$$
$$\times v^{\alpha}(\boldsymbol{q})_{s,s'} v^{\beta}(\boldsymbol{q})_{s',s} \frac{\frac{1}{2}[\Gamma(\boldsymbol{q})_s + \Gamma(\boldsymbol{q})_{s'}]}{[\omega(\boldsymbol{q})_s - \omega(\boldsymbol{q})_{s'}]^2 + \frac{1}{4}[\Gamma(\boldsymbol{q})_s + \Gamma(\boldsymbol{q})_{s'}]^2}.$$

$$(1)$$

$\kappa_P^{\alpha\beta}$ is not written out explicitly, as it corresponds to the well-known BTE expression for the thermal conductivity.[17,56,57] In the part of the equation describing the "coherences" conductivity, $\omega$, $C$, and $\Gamma$ represent the phonon frequencies, mode heat capacities, and linewidths for modes $s$ and $s'$ at the wave vector $\boldsymbol{q}$, while $v$ is the Wigner generalized group velocity operator[27] for cartesian directions $\alpha$ and $\beta$ that includes coupling between the modes $s$ and $s'$.

## Heat transport in polycrystalline acenes: comparing theory and experiment

We start the discussion by comparing our predictions with experimental data from literature. In the case of napthalene (2A), these are experimental results on compacted powder[67], while for pentacene (5A), experiments on thin films thermally evaporated on a Si/SiO$_2$ substrate are considered[68]. In both cases, samples were fabricated without control over the crystallite orientation and in the absence of a more in-depth structural characterization, we assume a polycrystalline texture for both experiments. To compare our simulations to these experiments, we will use the average of the trace of the conductivity tensor, $\kappa^{av}$, as an upper-bound estimator and the harmonic mean, $\kappa^{ha}$, as a lower-bound estimator for an isotropic thermal conductivity[69]:

$$\kappa^{av} = \frac{1}{3}(\kappa_{xx} + \kappa_{yy} + \kappa_{zz});$$
$$\kappa^{ha} = \left( \frac{\kappa_{xx}^{-1} + \kappa_{yy}^{-1} + \kappa_{zz}^{-1}}{3} \right)^{-1}.$$

$$(2)$$

A comparion between experimental and simulation data is provided in Fig. 3, where the displayed temperature ranges are adapted to the available experimental data. For the simulations, the total thermal conductivity, $\kappa_{tot}$ (black, solid line), the propagation conductivity, $\kappa_P$ (blue, dashed), and the tunnelling conductivity, $\kappa_C$ (red, dot-dashed), are shown. For these, the upper edges of the shaded areas correspond to $\kappa^{av}$ and the lower edges to $\kappa^{ha}$, while the lines (as averages between $\kappa^{av}$ and $\kappa^{ha}$) are guides to the eye and connecting the crosses that indicate the temperatures for which the WTE is explicitly evaluated.

For naphthalene, the experiments by Ueberreiter & Orthmann[67] were conducted with the hot wire method between 288 K and 340 K and are shown as orange circles in Fig. 3a. Results from the standard Peierls-Boltzmann transport equation (blue shaded range) clearly underestimate the measured thermal conductivities. In contrast, the WTE (gray shaded area) reproduces the measured thermal conductivities extremely well, especially up to ~310K. The discrepancy between the BTE and WTE results arises from the fact that $\kappa_C$ amounts to already 27% of $\kappa_P$ at 275 K and increases to even 37% at 350 K. This shows that for a quantitatively reliable description of heat transport in OSCs, it is crucial to include the tunneling conductivity. Another peculiarity is that, while in many crystals the thermal conductivity decreases with temperature roughly with $T^{-1}$ due to an increase of scattering[36], for 2A a much weaker temperature dependence is observed: in the simulations, the averaged thermal conductivity decreases with $T^{-0.46}$, which is very close to the $T^{-0.53}$ dependence obtained when fitting the experiments up to 302 K (see Fig. 3a). The reasons for the reduced dependence on temperature are that $\kappa_P$ decreases with only $T^{-0.77}$ and, even more importantly, that $\kappa_C$ distinctly increases with temperature. In passing we note that for pure Umklapp scattering of phonons one would expect an exponent of -1 only at temperature above the Debye temperature.[36] Hence, the observation that the negative exponent for $\kappa_P$ is smaller than 1 is attributed to the optical bands, that increasingly participate in the particle-like transport with rising temperatures.

A deviation from the above-mentioned exponent is observed in the experiments above ~310 K, where the experimental thermal conductivity decreases significantly stronger with temperature (namely with $T^{-1.63}$; see purple line in Fig. 3a). The authors of ref. 67 attribute this observation to the impact of vacancies, whose density substantially increases close to the melting point of naphthalene at 353 K (vertical dash-dotted line in Fig. 3a). This causes more scattering of phonons at defects. Additional influences of higher-order phonon scattering close to the melting point can also not be ruled out. As such effects are not considered in the simulations, the experimentally observed kink in the temperature evolution around 315 K cannot be reproduced. In passing we note the impact of boundary scattering for finite-sized crystal grains was not considered in our simulations for naphthalene, because the morphology of the used samples does not become apparent from the description in ref. 67. In naphthalene, however, the boundary scattering does not have a profound impact on the results presented in Fig. 3a, as is further discussed in Supplementary Section 4.1. There, it is shown that that a noticable decrease of the thermal conductivity occurs only for grain sizes below 100 nm. Finally, it should be mentioned that we are aware of a second set of temperature-dependent data for the thermal conductivity of 2A.[70] As those experiments were performed at elevated pressures, which were not considered in the MTP training, we refrain from explicitly discussing these data here.

For pentacene, the thermal conductivity of a polycrystaline thin film on a Si substrate was measured with frequency domain thermoreflectance by Epstein et al.[68] (orange square in Fig. 3b). For the samples in ref. 68 also the thin-film morphology was investigated by atomic force microscopy. In accordance with these experiments, boundary scattering for pentacene crystallites with a mean grain size of 256 nm was included in the evaluation of the WTE (see Methods section). On average, it reduces the thermal conductivity by around 7%. Results from the WTE are again in excellent quantitative agreement with the experimental data points over the entire temperature range. Notably, this excellent agreement is observed here between 75 K and 300 K, i.e., for a very wide temperature range.

Interestingly, the excellent agreement is also observed for the essentially temperature-invariant value of $\kappa$ close to 300 K. Splitting $\kappa$ into the propagation and tunneling contributions reveals that this is a consequence of an increase of inter-band phonon tunneling at elevated temperatures.

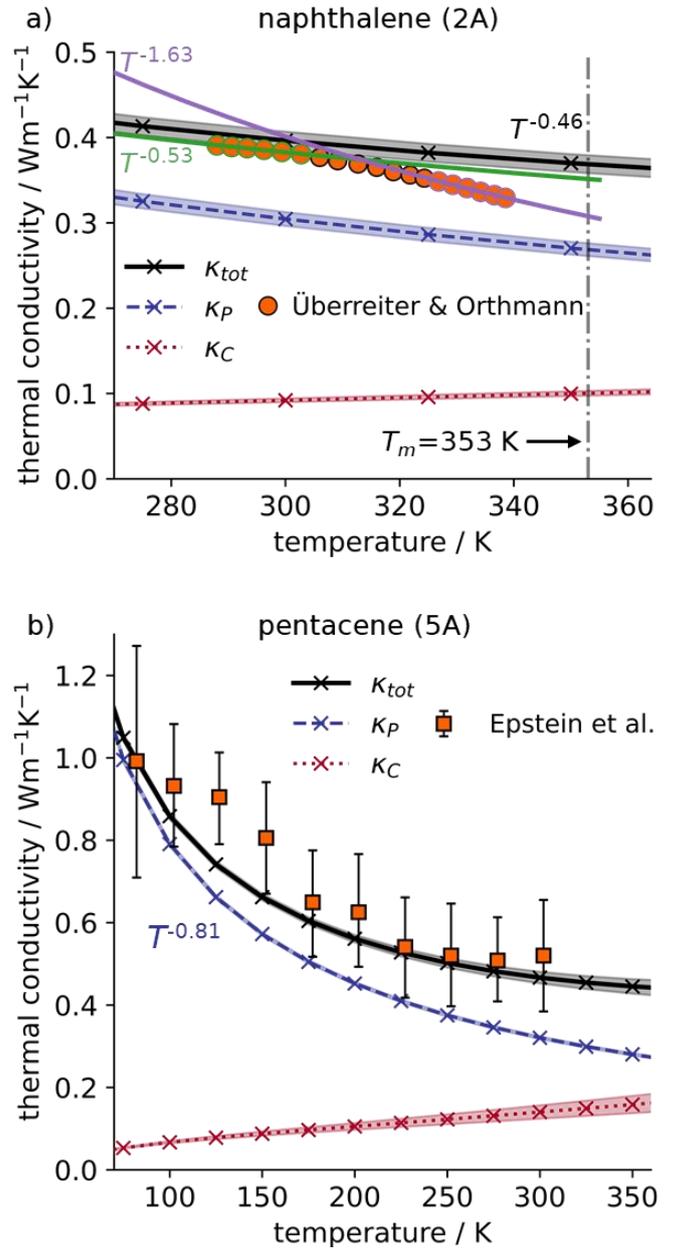

Fig. 3: Isotropic lattice thermal conductivity of naphthalene a) and pentacene b) as a function of temperature. The shaded areas indicate the isotropic thermal conductivities within the upper and lower bounds, as given by Eq. (2), while the lines show the averages of the upper and lower bounds and serve as a guide to the eye. Solid black lines refer to the total thermal conductivity, dashed blue lines denote the propagation contribution $\kappa_P$, and dotted red lines show the additional contributions due to phonon tunneling $\kappa_C$. Experimental results for naphthalene a) from ref. 67 are shown as orange circles and for pentacene b) from ref. 68 as orange squares. Furthermore, boundary scattering for a mean grain size of 256 nm (as determined in ref. 68) is accounted for in the calculation for pentacene. In panel a), two fitted lines for the first (green) and the last (purple) six experimental data points are drawn and the melting temperature of 353 K is indicated by a dash-dotted vertical line. In Supplementary Fig. 19, the data presented in panel a) are shown for a wider range from 70 K to 360 K.

This additional transport channel compensates the drop of the propagation contribution that arises from an increase of anharmonic (umklapp) scattering with temperature. This essentially complete cancellation of the temperature-dependent trends is not observed in 2A and a consequence of $\kappa_C$ being significantly larger in the longer acene. The compensation is in line with the observation for materials like perovskites[27,32], meteoritic tridymite[33], and Lanthanum-based materials for thermal barriers[27,34]: also in these materials, comparably complex phonon band structures with many overlapping bands result in strong $\kappa_C$ contributions.

## Heat transport in polycrystalline acenes: mode-resolved contributions

To rationalize the observed trends and to support the arguments provided so far, it is useful to analyze the contributions of individual phonon modes to the thermal conduction. An in-depth analysis of the (harmonic) phonons in the acenes, including their separation into inter- and intramolecular modes and an identification of the displacement types of each band, can be found in ref. 18. Here, we will primarily outline the overall situation, which is rather complex with overlapping acoustic and optical phonon bands. More information on the distinction between acoustic and optical bands in the case of complex phonon band structures is provided in Supplementary Section 6.

Intermolecular vibrations typically correspond to rigid body motions of entire molecules, i.e., they represent translations or rotations with respect to the other molecules and are, thus, directly affected by the molecular packing in the crystal. The associated eigenfrequencies are typically rather low (up to ~4THz) due to the weak intermolecular interaction strength combined with relatively large molecular masses. These modes form the acoustic and nine low-lying optical phonon bands. Moreover, in the acenes there are also several low-frequency optical bands formed by internal vibrations of the molecules. They typically correspond to distortions of the molecular backbones (e.g., bending vibrations) and the number of these vibrations increases with the size of the acene.[18] Most of these intramolecular modes form comparably flat bands with negligible group velocities. This implies a negligible contribution to the particle-like propagation conductivity. Still, their small inter-band spacings lead to sizable contributions to the wave-like tunneling conductivity.

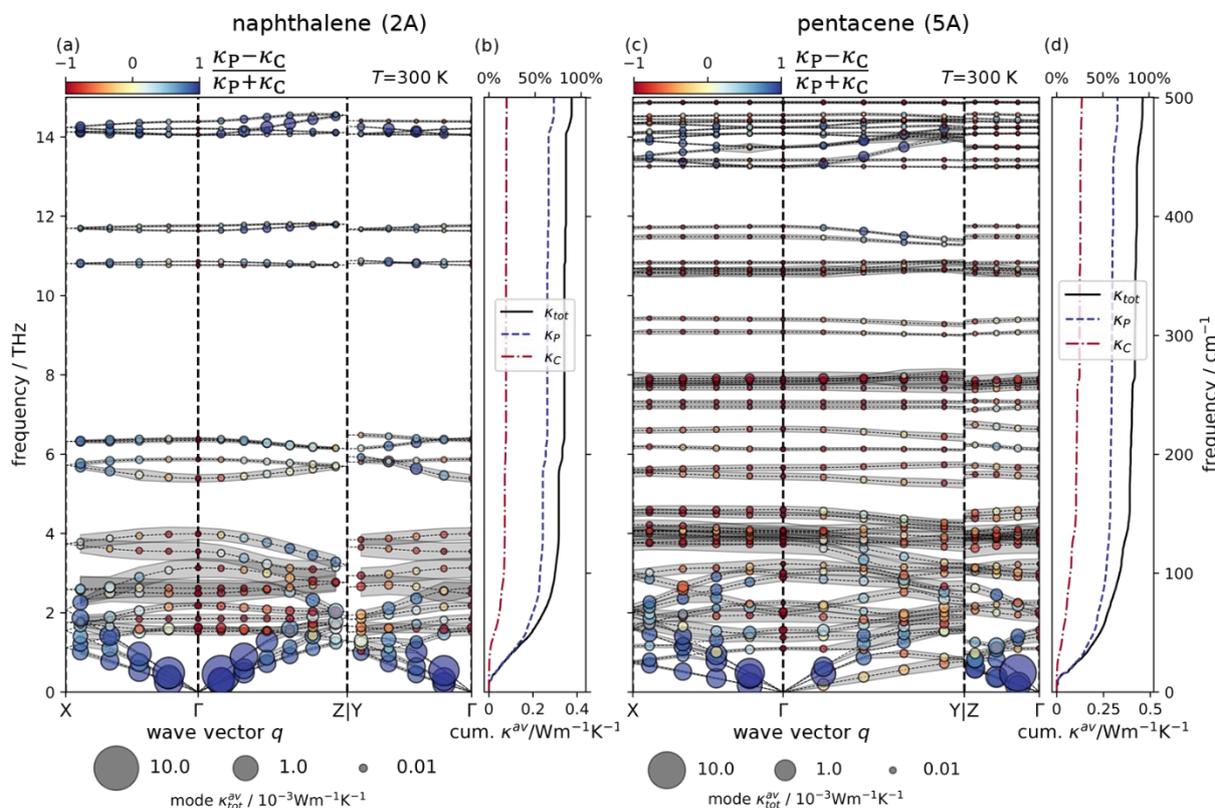

Fig. 4: Phonon band structure of a) naphthalene and c) pentacene (dotted black lines) with linewidths $\Gamma$ at T=300 K shown by the grey shaded areas (representing the full width at half maximum of the Lorentzian spectral functions). The mode contributions to $\kappa_{tot}^{av}$ are indicated by the diameters of the colored circles. The color indicates the dominant conduction mechanism (propagation or tunneling) according to the color bar on top of the panel. b,d) Cumulative mode contributions to $\kappa_{tot}^{av}$ as a function of the vibrational frequency at T=300 K. The solid black line represents the total thermal conductivity, the dashed blue line the propagation, and the dash-dotted red line the tunneling contributions.

The mode-resolved thermal conductivities in naphthalene and pentacene are shown in Fig. 4a and Fig. 4c. They are depicted as colored circles for selected points along the high-symmetry paths in reciprocal space. The diameters of the circles are scaled with the respective mode contribution to $\kappa_{tot}^{av}$ (see legend below the panel) and the colors illustrate the relative strength of propagation-type and tunneling-type conduction mechanisms at 300 K. A legend of the color scale and the definition of the encoded quantity are provided in the top part of Fig. 4. As complementary information, Fig. 4b and Fig. 4d illustrate the cumulative thermal conductivity, i.e., the overall thermal conductivities due to all phonons below a certain frequency. Again, particle-like and wave-like contributions are shown separately. Overall, the contributions of modes with frequencies above 4 THz are comparably small for both acenes, although, around 6 THz and 14 THz additional modes with non-negligible contributions are observed (see also discussion below).

Regarding the mechanism by which the various phonon modes contribute to the thermal conductivity, one finds that the acoustic phonons transport heat primarily in a particle-like manner, i.e., $\kappa_P$ dominates, as indicated by the blue shadings of the circles in Fig. 4a and Fig. 4c. This becomes apparent also from the cumulative thermal conductivity in Fig. 4b and Fig. 4d, where for modes up to 1 THz $\kappa_{tot}^{av}$ is entirely dominated by $\kappa_P^{av}$ and has already reached ca. 50% of its final value. The very high particle-like contribution for the acoustic phonons can be explained by the large associated group velocities, by the large mode contributions to the heat capacity resulting from the low frequency of the phonons, and by the comparably large phonon lifetimes, evidenced by the rather small anharmonic linewidths (gray shadings) in Fig. 4a and Fig. 4c.

For the optical modes between 1.5 THz and 4 THz the situation is reversed: phonon group velocities are rather small, while the anharmonic broadening is significantly larger than for the acoustic bands. Additional comparisons to experimentally-determined Raman spectra linewidths as functions of temperature are provided in Supplementary Section 3. In this frequency region, individual mode contributions to the thermal conductivity are small and in many cases phonon-tunneling dominates (red shading). The latter is a consequence of the significant overlap of the broadened band linewidths in that frequency region. Consequently, the cumulative of $\kappa_C^{av}$ in that frequency region increases somewhat faster than the cumulative of $\kappa_P^{av}$, even in absolute numbers. At higher frequencies, there are, on the one hand, phonon band-gaps and, on the other hand, certain phonon bands, which primarily contribute via propagation. This is a consequence of the relatively large group velocities and rather large lifetimes of these modes.

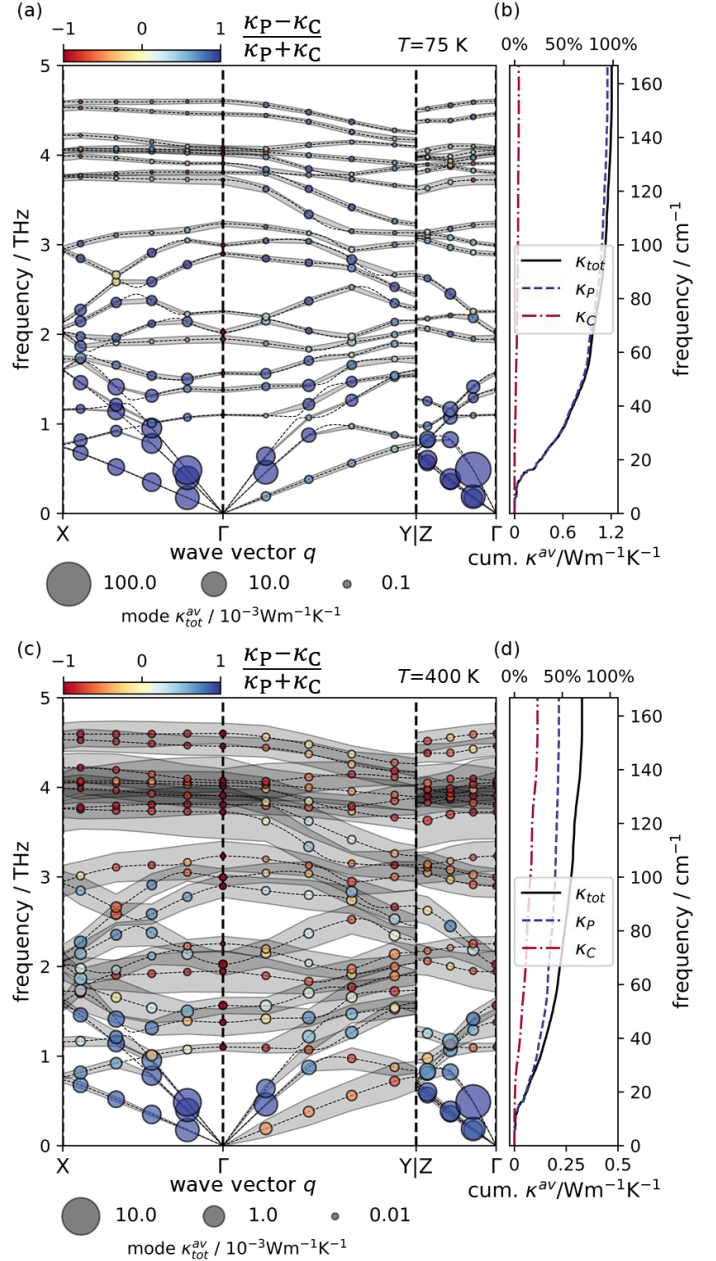

Fig. 5: Phonon band structure of pentacene (dotted black lines) with anharmonic linewidths at (a) T=75 K and (c) T= 400 K shown by the grey shaded areas (representing the full width at half maximum of the Lorentzian spectral functions). The mode contributions to $\kappa_{tot}^{av}$ are indicated by the diameters of the colored circles. Please note that the scales for the circle diameters differ by one order of magnitude between the two panels. The color of the circle reflects the dominant conduction mechanism (particle-like or phonon tunneling) according to the color bar on top of the panel. (b,d) Cumulative mode contributions to $\kappa_{tot}^{av}$ as a function of the vibrational frequency at T=300 K. The solid black line represents the total thermal conductivity, the dashed blue line the propagation, and the dash-dotted red line the tunneling contributions.

Their lifetimes are attributed to reduced three-phonon scattering rates due to the associated selection rules in conjunction with the gaps. However, a detailed analysis of such aspects goes beyond the scope of the present publication.

To understand the temperature dependence of the coexisting propagation and tunneling conductivities, Fig. 5 compares the linewidth broadening and the mode contributions for pentacene at 75 K (Fig. 5a) and at 400 K (Fig. 5c). The focus here is on the modes below 5 THz, as these modes yield essentially 100% of the thermal conductivity at 75 K and still ~70 % at 400 K. The corresponding cumulative conductivities are contained in Fig. 5b and Fig. 5d. At 75 K, the thermal conductivity is almost entirely the results of particle-like phonon propagation. This is a consequence of very large phonon lifetimes due to the sharply reduced probability of (Umklapp) scattering processes at low temperatures. As a result of these well-defined bands, the phonon tunneling is diminished. In contrast, the massive reduction of phonon lifetimes at 400 K reduces particle-like transport and $\kappa_P^{av}$ below 5 THz accounts for only 47% of the overall thermal conductivity. Moreover, the substantial overlapping of phonon bands due to the broadening (see Fig. 5c) boosts phonon tunneling. Thus, the tunneling conduction dominates for most of the dense-lying optical phonons. Interestingly, the large line broadening at 400 K also causes the tunneling to become the dominant transport mechanism of one of the transversal-acoustic band along the $\Gamma Y$ path.

Additionally, at elevated temperatures, higher lying modes become occupied, increasing their relative relevance for the transport of heat. To quantify the discussed effects; at 75 K strong particle-like transport results in a thermal conductivity $\kappa_P^{av}$ of 1.1 Wm$^{-1}$K$^{-1}$. It decreases with temperature to a value of 0.22 Wm$^{-1}$K$^{-1}$ at 400 K. The broadening of the phonon bands at that temperature is, however, so large, that $\kappa_C^{av}$ has in turn reached a value of 0.11 Wm$^{-1}$K$^{-1}$ (up from 0.05 Wm$^{-1}$K$^{-1}$ at 75 K). Thus, the overall drop of the thermal conductivity between 75 K and 400 K is markedly reduced, with a close to temperature-invariant value of $\kappa_{tot}^{av}$ above 250 K (see Fig. 3b).

# Origins of the anisotropic thermal conduction in acenes

So far, we have focused on isotropic thermal conductivities, $\kappa^{av}$, primarily to allow a direct comparison to experiments performed on polycrystalline samples. Molecular crystals like acenes are, however, anisotropic. This is, for example, well established for the charge carrier mobility of pentacene,[71,72] raising the question, to what extent this anisotropy prevails also for the thermal conductivity.

Therefore, what follows is an analysis of the thermal conductivities along the crystallographic directions that are parallel to the lattice vectors. In the above discussion, we focused on 2A and 5A, as for these materials experimental data were available. We now include also the intermediate length acenes anthracene (3A) and tetracene (4A) in the comparison and note that, to the best of our knowledge, no direction-dependent heat transport measurements exist for any of the acenes. These systems are either monoclinic or triclinic, so their thermal conductivity tensors are composed of either four or six independent elements. To directly compare structures with different crystal symmetries, we calculate scalar thermal conductivities along specific directions. Generally, we compute the scalar conductivity $\kappa^d$ in an arbitrary direction $\boldsymbol{d}$ from the full conductivity tensor $\kappa$ as follows:

$$\kappa^d = \frac{\boldsymbol{d}^T \kappa \, \boldsymbol{d}}{|\boldsymbol{d}|^2}. \qquad (3)$$

This quantity describes the component of the heat flux in direction $\boldsymbol{d}$ due to a temperature gradient in the same direction. In the following, this equation will be applied to obtain the conductivities in the crystallographic [100], [010], and [001] directions and within the crystallographic (010) and (001) planes. These conductivities will be labeled accordingly (e.g., $\kappa_P^{[100]}$). Directions [100] and [010] are parallel to the (001) plane in which the acenes are herringbone stacked. The [001] direction is (close-to) parallel to the long molecular axes.

The respective lattice thermal conductivities $\kappa_{tot}$ (solid) and their contributions $\kappa_P$ (dashed), and $\kappa_C$ (dash-dotted) are shown in Fig. 6 for 2A (purple), 3A (magenta), 4A (blue), and 5A (orange). The different panels show $\kappa^{[100]}$ (a), $\kappa^{[010]}$ (b), and $\kappa^{[001]}$ (c). For the sake of completeness, the thermal conductivity tensors at 300 K are listed in Supplementary Section 5 for each material.

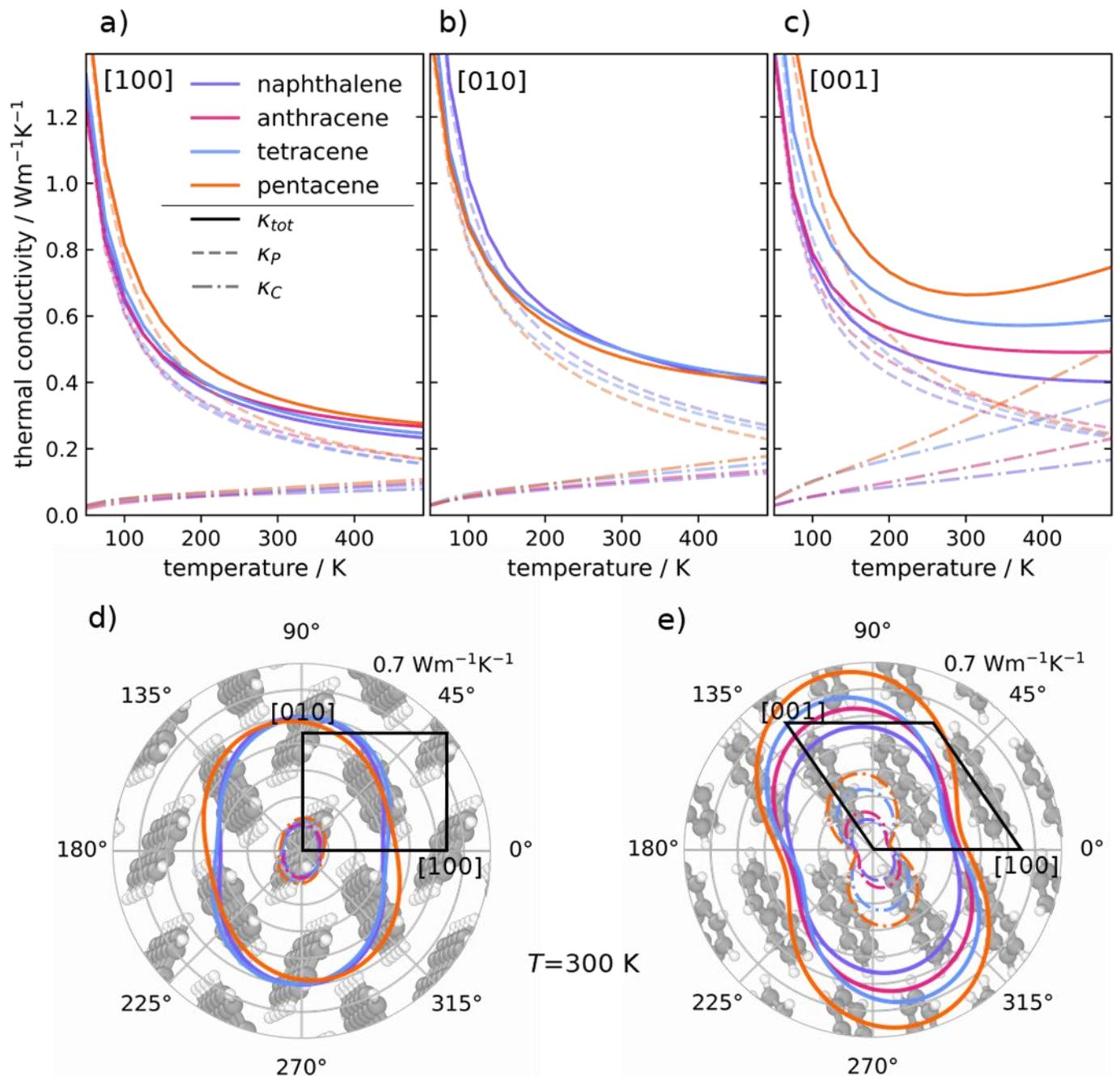

*Fig. 6: Anisotropic lattice thermal conductivities of naphthalene (purple), anthracene (magenta), tetracene (blue), and pentacene (orange) as a function of temperature. Solid lines show the total thermal conductivity $\kappa_{tot}$, dashed lines the propagation contribution $\kappa_P$ and dash-dotted lines the tunneling contribution $\kappa_C$. The anisotropic conductivities are presented for the crystal directions [100] in panel a), for [010] in panel b), and for [001] in panel c). Polar plots of $\kappa_{tot}$ (solid lines) and $\kappa_C$ (dash-dotted lines) at T=300 K within the (001)-plane and within the (010)-plane are shown in panel d) and e). The magnitude of $\kappa_P$ can be deduced from the difference between $\kappa_{tot}$ and $\kappa_C$. The polar plots are superimposed on the crystal structures of pentacene (d) and naphthalene (e) for the sake of illustration. The respective crystal directions are labeled. The gray rings indicate increments of 0.1 Wm$^{-1}$K$^{-1}$. In panels b) and d), $\kappa_{tot}$ and $\kappa_P$ of anthracene are omitted for reasons mentioned in the main text.*

From the presented data we can deduce relevant structure-property relationships in the acenes: For the [100] direction (see Fig. 6a), the decrease of the thermal conductivities with temperature is very similar for all systems. This also applies to each of the two conduction channels ($\kappa_P$ and $\kappa_C$). Likewise, in the [010] direction there is also no pronounced difference between the acenes. In passing we note that for 3A, $\kappa_P^{[010]}$ and $\kappa_{tot}^{[010]}$ are not shown, since the standard broadening schemes for the calculation of the phonon linewidths do not show clear computational convergence (as discussed in Supplementary Section 2.6). This suggests that the completed-collision limit considered in the derivation of the Wigner transport equation might not be sufficient to capture all the physics. Thus, in this case, spectral functions[66,73,74] or collisional-broadening effects[75] might have to be considered.

On average, the thermal conductivities (as well as the contributions of both transport mechanisms) are by around 33% smaller along [100] than along [010] for all acenes. Thus, plotting the directional dependence of $\kappa_{tot}$ at 300 K within the (001) plane in Fig. 6d reveals a distinct anisotropy, where the curves of 2A and 4A display maxima parallel to the [010] and minima parallel to the [100] direction. In 5A, the direction of maximum thermal conductivity clearly deviates from the [010] direction, which is attributed to its triclinic symmetry. This is the primary reasons, why $\kappa_{tot}$ of 5A below 300 K is larger in the [100] and smaller in the [010] direction, compared to the other acenes. This trend prevails, even though, as shown in Supplementary Fig. 23, the direction of the maximum thermal conductivity in the polar plots varies with temperature. Another relevant aspect is, that $\kappa^{[100]}$ and $\kappa^{[010]}$ are dominated by the particle-like conduction mechanism and that the increase of the tunneling contributions with the size of the acenes is barely noticeable.

While thermal conductivities within the (001) plane are almost material-independent, this is not the case for the [001] crystal direction. In that direction, only $\kappa_P^{[001]}$ is essentially the same for all systems. In contrast, $\kappa_C^{[001]}$ significantly increases with the molecular length, especially at higher temperatures, where the thermal conductivity clearly scales with the size of the molecules. This leads to a substantial increase of the total thermal conductivity in [001] direction. For 4A and 5A, $\kappa_C^{[001]}$ even surpasses $\kappa_P^{[001]}$ around 350 K and dominates at higher temperatures. This results in $\kappa_{tot}^{[001]}$, beyond 375 K for 4A and beyond 300 K for 5A, eventually even rising with $T$. Notably, such a behavior is typically not observed in crystalline materials close to room temperature, but occurs in amorphous solids.[76]

The increase of $\kappa_C$ with molecular length also causes a pronounced growth in the anisotropy of $\kappa_{tot}$, as is illustrated for the (010) plane in Fig. 6e for a temperature of 300 K. The maxima of $\kappa_{tot}$ within the (010) plane occur close-to parallel to the long molecular axes and the minima are essentially parallel to the [100] direction. Consistent with the above discussion, these minima hardly vary between the different crystals while the maxima become successively larger for longer molecules.

To understand, why $\kappa_C^{[001]}$ increases with molecular length while, e.g., $\kappa_C^{[100]}$ stays nearly constant, we analyze their cumulative (spectral) contributions at $T$=300 K, which are plotted in Fig. 7a for [100] and in Fig. 7b for [001]. In contrast to the corresponding plots in Figs. 4 and 5, here the frequency axis is plotted on a log scale to accentuate the narrow spectral region below 3-4 THz (green shaded area). For both, the [100] and [001] directions we see an inflection point around 3 THz. Up to this frequency, mostly intermolecular, rigid-body vibrations occur.[18] The cumulative contributions of these modes to $\kappa_C^{[001]}$ are identical for each acene, and also the values of $\kappa_C^{[001]}$ display only minor variations. From that we can deduce that the twelve intermolecular modes in the different acenes contribute rather equally to $\kappa_C$, irrespective of the direction.

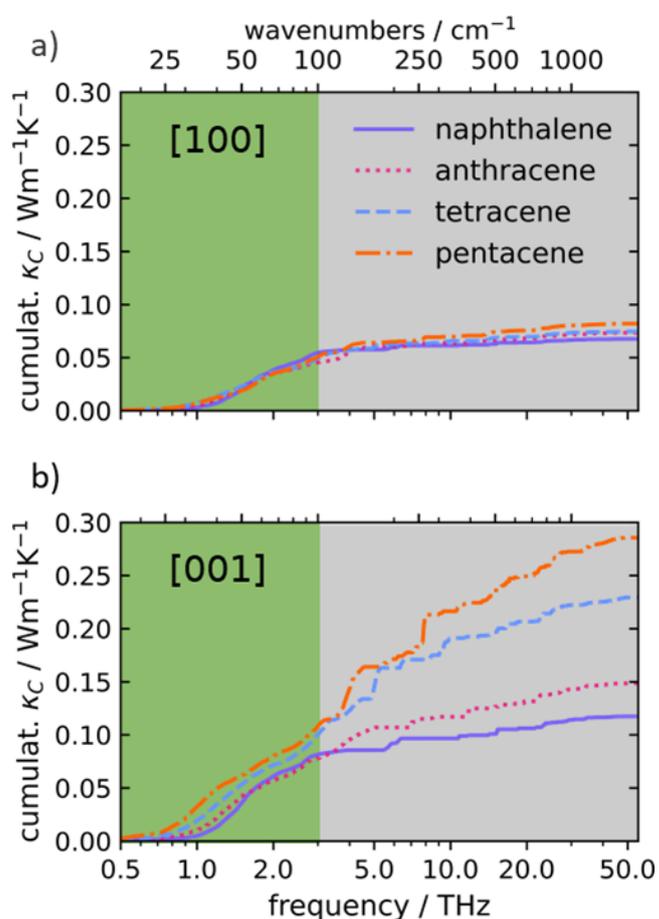

*Fig. 7: Cumulative mode contributions to the tunneling conductivities $\kappa_C^{[100]}$ in a) and $\kappa_C^{[001]}$ in b) at T=300 K as a function of vibrational frequency (plotted on a log scale). The region of (mostly) intermolecular vibrations ("rigid body modes") below 3 THz is shaded in green while the region of intramolecular vibrations is shaded in gray.*

Beyond 4.25 THz the modes correspond to intramolecular, backbone-distorting vibrations.[18] In the [100] direction, higher-frequency phonons contribute less than those below 3 THz and the variation between the acenes is only 0.02 Wm⁻¹K⁻¹ (Fig. 7a). For the cumulative of $\kappa_C^{[001]}$ an entirely different behavior is observed in Fig. 7b, where for the longer acenes the cumulative mode contributions grow at a sizable rate. Therefore, the stark difference in $\kappa_C^{[001]}$ between the acenes can be primarily attributed to the contributions of the intramolecular modes (grey shaded area). This is consistent with what becomes apparent from Fig. 4, namely that the vast majority of the additional optical bands in the longer acenes are found above 3 to 4 THz and that these bands significantly contribute to the thermal transport via phonon tunneling.

The observation that the contribution of these modes is particularly pronounced in the [001] direction can be correlated with the fact that in this direction the ratio of covalent bonds to van der Waals bonds distinctly increases for longer acenes, which is apparently favorable for thermal transport due to the stronger interatomic coupling. On more formal grounds, one can argue that the stronger tunneling is a consequence of an increased number of optical bands (i.e., more phonon states)[18] in the longer acenes. Nevertheless, in directions perpendicular to the long molecular axis, e.g. parallel to the (100) and (010) planes, the effect of the larger number of bands is compensated by the increased cross-section of the crystal. In contrast, the face of the unit cell parallel to the (001) plane is largely independent of the length of the acene molecule, hence why the larger of bands leads to a higher tunneling conductivity.

From an intuitive viewpoint, proceeding from 2A to 5A can be interpreted as transitioning from a "simple molecular crystal" to a "complex covalent crystal". In this analogy, the simple molecular crystal features a small primitive cell, containing small molecules, with a short periodicity length scale and high crystalline order. In such a system, the dispersion of the low-frequency phonons is determined by the weak intermolecular forces. Therefore, the group velocities of these bands and subsequently, in reference to Eq. (1), their velocity-operator elements would generally be low. This in combination with the fact that the small number of atoms leads to few and well-separated optical bands, results in weak tunneling. The complex covalent crystal, instead, is made up of primitive cells containing large molecules. The phonon dispersions are determined by strong intramolecular covalent forces, which imply large velocity-operator elements along the molecular backbone direction. The large number of atoms in the primitive cell implies a large number of closely spaced phonon bands. Facilitated by the large velocity-operator elements, the narrow band spacings allow strong wave-like tunneling along the molecular backbone direction.

## Discussion

In this study, heat transport in prototypical crystalline organic semiconductors is investigated at an unprecedented level of detail. The subject of this work is the series of acenes (from naphthalene to pentacene) where the size of the molecular building blocks systematically increases without significantly changing the intermolecular arrangement. Interestingly, due to their structural complexity and strong anharmonicity, these systems cannot be quantitatively described using the standard Peierls-Boltzmann description based on particle-like phonon propagation and scattering. To fully capture the relevant conduction mechanisms, it is essential to also account for wave-like phonon tunneling processes. This can be done within the framework provided by the Wigner transport equation (WTE).[27]

To solve the WTE, one needs to calculate anharmonic force constants. For the monoclinic or triclinic systems at hand this is not a trivial task, which is further complicated by the large number of atoms per unit cell. Performing the required single-point calculations with DFT would be prohibitively expensive. Therefore, we trained system-specific and highly accurate Moment Tensor Potentials[48,77] (MTP) on a few hundred DFT calculated reference structures.[47] These reference configurations were obtained by an active-learning approach implemented in the VASP code, which samples the relevant configuration space very efficiently. After validating their efficacy in describing the vibrations within the acenes, the MTPs were used to calculate (an)harmonic force constants from small atomic displacements.[55]

As illustrated in Fig. 3, when combining the WTE with the above-mentioned MTPs, we obtain an excellent quantitative agreement of the thermal conductivity between modelling and experiments[67,68]. Notably, the opposing temperature dependencies of the propagation and tunneling contributions explain the gradual decrease of the overall conductivity in naphthalene and the essentially temperature-independent thermal conductivity in pentacene around room temperature that are observed in experiments.

Analyzing the contributions of individual phonon modes at 300 K (Fig. 4), one finds that heat is mostly (>50%) transported via particle-like scattering of acoustic phonons, while the optical phonons contribute via phonon tunneling, providing between 15-20% of the total thermal conductivity already at room temperature.

Regarding the anisotropy of the thermal conductivities, we find that within the plane in which the molecules form a herringbone pattern, the total thermal conductivities of the different acenes are essentially identical over the entire considered temperature range (Fig. 6). In contrast, in directions close-to parallel to the long molecular axes, the tunneling conductivity increases with the size of the molecules while the propagation conductivity remains system independent. Especially for the longer acenes, this results in heat transport being more favorable along the long molecular axis than within the herringbone plane. This is in sharp contrast to what is generally observed for the electrical conductivity of organic semiconductors, which is typically at a maximum within the herringbone plane.[71,72,78,79] For pentacene and tetracene, the strong increase of $\kappa_C$ and the opposing temperature dependencies of $\kappa_P$ and $\kappa_C$ result in another peculiarity: around 375 K in tetracene and around 300 K in pentacene, the total thermal conductivity along the [001] direction displays a profound minimum before increasing again at higher temperatures. This anomalous behavior deviates from the saturating or weakly decreasing trend around room temperature normally observed in complex crystals[34] or glasses[80]. It is driven by anharmonicity and therefore it is fundamentally different from the one caused by the onset of radiative transport at extremely high temperatures.[81]

In conclusion, the data shown here illustrate that (i) system-specific machine-learned MTPs are capable of accurately simulating thermal conductivities of complex materials like organic semiconductors also when explicitly describe heat transport via phonons. (ii) This, however, works only, if one goes beyond a mere particle-like description of phonon transport in the framework of the Boltzmann transport equation and takes phonon tunneling into account, which can be done within a generalized heat transport model derived from the Wigner transport equation.[27] (iii) In fact, the in-depth analysis of the structure-property relations for heat transport in acenes reveals that phonon tunneling is the primary factor responsible for most of the peculiarities observed in these materials. It causes the weak (in naphthalene) or vanishing (in pentacene) temperature dependence of the average thermal conductivity around room temperature and it gives rise to an anisotropy of the thermal conductivity that is opposite to that observed for the electrical conductivity. Furthermore, it causes the pronounced dependence of $\kappa_{tot}^{[001]}$ on the length of the acene molecules, which for some materials even results in a pronounced minimum in the evolution of the thermal conductivity with temperature. We are confident that this work lays the foundation for understanding heat transport in organic semiconductors and related materials from a reciprocal-space perspective via the propagation and tunneling of phonons.

## Methods

### Machine-learned interatomic potentials

Training data to parametrize the machine-learned interatomic potential used in this work consist of energies, forces, and stresses calculated with dispersion-corrected DFT for specific configurations. These configurations are calculated along the trajectories of molecular dynamics (MD) simulations conducted in the Vienna ab-initio simulation package (VASP; version 6.4.1).[82–84] To efficiently sample the range of possible configurations of the molecular crystals, these MD trajectories are calculated using the active-learning routine implemented in VASP[85,86]. In short, during active learning a machine-learned force-field (ML-FF) is trained on the ab-initio data acquired during the first few timesteps. At each further timestep, the trajectory is updated by the hitherto parametrized ML-FF as long as the uncertainties of the resulting forces (estimated via Bayesian errors) do not exceed a threshold. Otherwise, the ML-FF prediction is discarded and the next timestep is calculated using DFT. This provides a new DFT reference configuration, which is eventually used for an on-the-fly update of the ML-FF. In this way, only a few hundred "most important" configurations are automatically selected for DFT calculations, even though typical trajectories comprise several thousand time steps (see below).

In a second step, the properties of the DFT-calculated reference configurations are used to parametrize state-of-the-art Moment Tensor Potentials (MTP)[48,77], in analogy to the procedure described in previous work.[47,55] Advantage of the MTPs are that they can be highly accurate, only need a manageable amount of training data, and that they can be directly interfaced[87] with LAMMPS[88], which allows for particularly efficient calculations of anharmonic force constants. What follows are the details for both the active-learning in VASP and the parametrization of the MTPs.

### Details of the active-learning MD

Active learning simulations were performed within an $NpT$ ensemble to allow changes in cell shape and volume, which in molecular crystals can especially result variations of stacking distances, molecular rotations, and the relative arrangements of the molecules. This expands the explored configuration space. In the simulations, we applied a temperature ramp and chose settings that are expected to be particularly suitable for eventually describing the relatively small displacements considered when calculating phonons from finite differences.[55] We chose a relatively low starting temperature of 50 K and steadily heated the systems to a moderate temperature of 200 K within 15,000 steps and a time step of 0.5 fs, applying a Langevin thermostat.[89] Notably, the final temperature is by a factor of around three smaller than in most previous studies[90] involving VASP active learning. This is because in the present case, the aim is to generate potentials that allows to calculate (an)harmonic interatomic force constants from small atomic displacements rather than parametrizing a force field for performing MD simulations. As shown in previous work,[55] for the intended task a higher level of accuracy is achieved, when parametrizing the MTPs on structures obtained during MD runs at lower temperatures. In the active learning, the initial force error threshold for triggering ab initio calculations was set to 2 meV/Å to enforce sufficient ab-initio steps in the early stages of the run. It is worth mentioning that in some cases the active learning routine would only calculate the very first time step with DFT, which resulted in unstable ML-FFs. In such a case, we ensured that the initial training data contained enough DFT configurations, by simulating a short trajectory of 50 steps with a stricter threshold that was fixed ("ML_ICRITERIA=0") to 0.1 meV/Å prior to the actual active learning. Other than that, the default values for the descriptors and machine-learning hyperparameters of VASP version 6.4.1. were used. Prior to the MD simulations, atomic positions within the primitive unit cells were optimized, as suggested in the VASP manual.[91] In the active learning runs, (1,2,1) or (2,1,1) super cells were used, resulting in 72 to 144 atoms in the supercells of the various acenes (see Supplementary Table 3). All ab-initio calculations were performed with VASP 6.4.1, employing the PBE exchange functional,[92] the D3[93] dispersion correction with Becke-Johnson damping by Grimme[94] and the standard projector-augmented wave pseudo-potentials.[95] The energy cutoff for the basis sets were set to 800 eV or 1050 eV depending on the type of simulations, as motivated in Supplementary Section 2.1. In that section, also further details on the applied settings (Supplementary Table 3) and convergence tests can be found.

## Parametrizing Moment Tensor Potentials

Moment tensor potentials (MTP)[48] were parametrized on the VASP ab-initio training data using the MLIP-2 package.[77,96] Preliminary tests showed that MTPs with the most sophisticated analytical structure provided by MLIP-2 (namely, level 28 MTPs) are best suited for the task of calculating accurate forces for finite differences approaches. This assessment was based on the observation that level 28 potentials provide the highest accuracy in terms of reproducing DFT-calculated phonon band-structures (see Results section), while the associated computational demand for calculating atomic forces of displaced structures remained essentially negligible. In the case of MTPs, "level" refers to a parameter that in combination with the chosen number of basis functions (which was here set to 20) determines the speed and accuracy of the MTP. While the maximum cutoff distance for the interactions was left at the default value of 5 Å, the minimum was reduced to be below the smallest occurring pair distance in the respective training data and a value of 0.95 Å was chosen for all four systems. As the initialization in the MLIP-2 training employs random parameters, the resulting MTPs have a stochastic nature. Thus, we trained three randomly initialized MTPs and analyzed their performance on a separate validation set of DFT configurations (details in Supplementary Section 2.2.1). The performance differences in all cases were very minor and much smaller than when using lower level MTPs, which some of us applied in previous studies for performing MD simulations on related materials.[47,55] All three independently parametrized MTPs produced energies, forces, and stresses with a close-to perfect correlation to the validation set (see Supplementary Fig. 3). Eventually, the MTP with the smallest force validation error (see Table 1) was also the best at predicting phonon frequencies (see Results section) and was then picked as the default MTP for a specific acene.

## Calculating force constants and phonon band structures

Second-order force constants were calculated from the atomic forces of displaced super cells with the finite-difference method[97] implemented in Phonopy and its python modules.[98–100] An atomic displacement amplitude of 0.01 Å was chosen. Prior to calculating the atomic forces, the ionic positions were relaxed with a force difference threshold of $10^{-8}$ eVÅ$^{-1}$. Both, the relaxation and the single-point force calculations were performed with an in-house routine utilizing MTPs via the MLIP LAMMPS interface[87] and the python modules of LAMMPS.[88] To allow a direct comparison between DFT- and MTP-based phonon frequencies, the same DFT-relaxed lattice parameters as in ref. 18 were used also in the MTP calculations. However, we note that relaxing the cells with MTPs results in virtually identical structures, as shown in Supplementary Table 2. The third-order force constants were obtained applying a similar strategy. The displaced structures were generated with Phono3py[57,100] and the forces on the atoms were calculated with the MTPs. From the atomic forces, the third-order force constants were then obtained with Phono3py accounting for translational and index-exchange symmetries. The atomic displacement amplitude used for calculating third-order force constants was set to 0.03 Å for all four acenes based on testing the impact of the chosen displacement on the thermal conductivity of naphthalene (Supplementary Fig. 8). Notably, second-order force constants calculated with an amplitude of 0.03 Å were potentially beyond the harmonic regime and resulted in minor frequency shifts for some acenes, which is why we used two different displacement amplitudes. Tests for naphthalene also suggested, that third-order force constants were converged for smaller supercell dimensions than those required for converging the second-order force constants (Supplementary Fig. 9). Hence, we saw no need to extend the supercells for calculating third-order force constants and we employed the same supercells that had already been used for calculating harmonic phonon properties.[18,46] The super cell dimensions and numbers of atoms are summarized in Table 2. We also want to stress that these supercells are larger than the ones used in the active learning listed in Supplementary Table 3.

Table 2: System-specific supercell dimensions (DIM), number of atoms in the supercells ($N_{atom}$), number of displaced structures for obtaining third-order force constants ($N_{disp}$), and q-meshes used for calculating phonon lifetimes and evaluating the Wigner transport equation.

| System | DIM | $N_{atom}$ | $N_{disp}$ | q-mesh |
|--------|-----|------------|------------|--------|
| 2A | (2,3,2) | 432 | 140022 | 9×13×9 |
| 3A | (2,3,2) | 578 | 248904 | 10×10×8 |
| 4A | (2,2,2) | 480 | 518580 | 10×10×8 |
| 5A | (2,2,2) | 576 | 746712 | 9×9×7 |

## Computing lattice thermal conductivities

Thermal conductivities were calculated employing the Wigner transport equation as implemented in phono3py[57,100] (version 2.7.0) by the authors of refs.27,28. Phonon scattering rates were calculated with the tetrahedron method for Brillouin zone integration.[101,102] For each material, the number of sampled reciprocal space points (q-mesh) was individually tested (see Supplementary Section 2.4) and the converged meshes are listed in Table 2. All presented thermal conductivities were obtained within the relaxation time approximation (RTA), after confirming its validity for the materials at hand by comparing it to the direct solution of the linearized Boltzmann transport equation (LBTE)[103] for naphthalene. While directly solving the LBTE massively increases the computational cost, the differences in thermal conductivity between both methods were negligible (see Supplementary Fig. 7). This is consistent with previous studies on low-thermal conductivity solids.[27,104] Further, we made sure that the computed phonon lifetimes are not overdamped and above the

Ioffe-Regel limit[105] (see Supplementary Section 2.5). To account for boundary scattering in finite-sized crystal grains when comparing the WTE results with experiments (see Fig. 3), an additional scattering rate is included within phono3py ("--boundary-mfp [$L$]"). This scattering rate is given by $v_g/L$, with $v_g$ representing the group velocity of a given mode and $L$ denoting the chosen mean free path between the grain boundaries. It was added to the phonon-phonon scattering rates via Matthiessen's rule.

## Data availability

Datasets generated and/or analyzed during the current study are available at the repository of Graz University of Technology; https://doi.org/10.3217/t9czy-wjy83

## Code availability

VASP can be acquired from the VASP Software GmbH (www.vasp.at/); LAMMPS is available at www.lammps.org; MLIP is available at mlip.skoltech.ru/download; the lammps-mlip interface is available at gitlab.com/ashapeev/interface-lammps-mlip-2; Phonopy is available at phonopy.github.io/phonopy; Phono3py is available at https://phonopy.github.io/phono3py

# Acknowledgments

We acknowledge Tomas Kamencek and his work in laying the foundation for this publication. We also would like to thank Lukas Hörmann for stimulating discussions and the lecturers, tutors, and participants of the TDEP2023 summer school in Linköping for their educational insights. Further, we thank Ferenc Karsai for helping with some technical details of the VASP active-learning.

This research was funded by the Austrian Science Fund (FWF) [primarily Grant-DOI: 10.55776/P33903 and in part also 10.55776/P36129]. For the purpose of open access, the authors have applied a CC-BY public copyright license to any author accepted manuscript version arising from this submission. We also acknowledge the Graz University of Technology for support through the Lead Project Porous Materials @ Work for Sustainability (LP-03). Computational results have been obtained using the Vienna Scientific Cluster, VSC-5.

M. S. acknowledges support from: (i) Gonville and Caius College; (ii) the UK National Supercomputing Service ARCHER2, for which access was obtained via the UKCP consortium and funded by EPSRC [EP/X035891/1]; (iii) the Kelvin2 HPC platform at the NI-HPC Centre (funded by EPSRC and jointly managed by Queen's University Belfast and Ulster University).

# Author information

**Contributions:** Conceptualization, L.L., E.Z.; methodology, L.L., L.R., S.W., and M.S.; software, L.L., L.R., S.W., and M.S.; validation, L.L., E.Z.; formal analysis, L.L.; investigation, L.L., L.R., S.W., M.S., E.Z.; resources, E.Z.; data curation, L.L.; writing—original draft preparation, L.L., E.Z.; writing—review and editing, E.Z., L.L., L.R., S.W., and M.S.; visualization, L.L.; supervision, E.Z.; project administration, E.Z., L.L.; funding acquisition, E.Z. All authors have read and agreed to the published version of the manuscript.

# Ethics declarations

**Competing interests**

The authors declare no competing interests.

# Heat transport in crystalline organic semiconductors: coexistence of phonon propagation and tunneling

Lukas Legenstein[1], Lukas Reicht[1], Sandro Wieser[1,2], Michele Simoncelli[3,4], and Egbert Zojer[1]

[1]Institute of Solid State Physics, Graz University of Technology, NAWI Graz, Petersgasse 16, 8010 Graz, Austria
[2]Institute of Materials Chemistry, TU Wien, Getreidemarkt 9, 1060 Wien, Austria
[3] Theory of Condensed Matter Group, Cavendish Laboratory, University of Cambridge (UK)
[4]Department of Applied Physics and Applied Mathematics, Columbia University, New York (USA)

## Table of Contents

# Supplementary Section 1: Crystal structures of the acenes

In Figure 1 of the main text, only the smallest and largest systems were shown. Supplementary Figure 1 contains the structures of all studied materials.

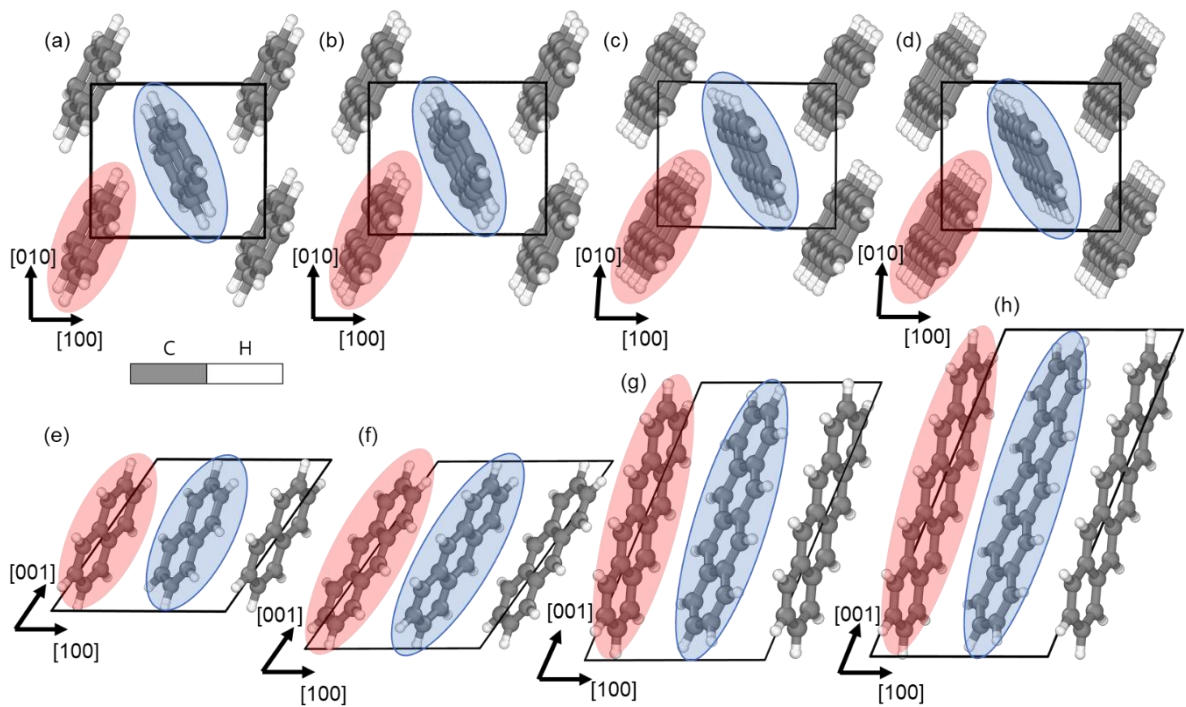

*Supplementary Figure 1: Crystal structures of the (oligo)acenes viewed along [001] (a-d) and along [010] (e-h) for naphthalene, anthracene, tetracene, and pentacene (left to right).*

The initial crystal structures of the stable, monoclinic and triclinic acene polymorphs studied in this work were obtained from the Cambridge Crystallographic Data Centre[1] and their identifiers are listed in Supplementary Table 1. Notably, for tetracene, lattice vectors $a_1$ and $a_2$ were switched compared to the database entry, to be consistent with the other acenes.

*Supplementary Table 1: Cambridge Crystallographic Data Centre (CCDC) identifiers of the considered acene structures together with the corresponding references.*

| acene | abbreviation | identifier | publication |
|---|---|---|---|
| naphthalene | 2A | NAPHTA31 | Capelli et al.[2] |
| anthracene | 3A | ANTCEN14 | Brock and Dunitz[3] |
| tetracene | 4A | TETCEN01 | Holmes et al.[4] |
| pentacene | 5A | PENCEN | Campbell et al.[5] |

The phonon band structures and lattice thermal conductivities presented in the main text have been calculated for the DFT-optimized lattice parameters from ref. 6. However, optimizing the unit cells with the respective system-specific Moment Tensor Potential (MTP) using a conjugate gradient method resulted in virtually the same lattice parameters. Both sets of lattice parameters (DFT and MTP) are presented in Supplementary Table 2, providing further reassurance regarding MTP accuracy, by showing that the trained MTP can properly describe the structure at and around the energetic equilibrium.

*Supplementary Table 2: Unit cell lattice parameters of the acenes optimized either with DFT (from ref. 6) or using the system-specific MTPs (this work). Note, that the presented unit cells for pentacene (5A) follow the convention of the experimentally determined cell by Campbell et al.[5] while the unit cells of tetracene (4A) represent reduced cells, based on Niggli's criteria[7] for space-group P-1.*

| acene | method | $a$ / Å | $b$ / Å | $c$ / Å | $\alpha$ / ° | $\beta$ / ° | $\gamma$ / ° | $V$ / Å³ |
|-------|--------|---------|---------|---------|-----------|----------|-----------|----------|
| 2A | DFT | 8.09 | 5.92 | 8.62 | 90.0 | 124.6 | 90.0 | 340.0 |
|    | MTP | 8.07 | 5.91 | 8.62 | 90.0 | 124.4 | 90.0 | 339.4 |
| 3A | DFT | 8.43 | 5.93 | 11.11 | 90.0 | 125.4 | 90.0 | 452.3 |
|    | MTP | 8.42 | 5.93 | 11.11 | 90.0 | 125.4 | 90.0 | 451.6 |
| 4A | DFT | 6.02 | 7.78 | 12.91 | 77.1 | 72.6 | 85.5 | 561.7 |
|    | MTP | 5.99 | 7.78 | 12.90 | 76.6 | 73.0 | 85.6 | 559.3 |
| 5A | DFT | 7.73 | 6.07 | 15.84 | 103.3 | 113.1 | 85.1 | 665.6 |
|    | MTP | 7.72 | 6.07 | 15.83 | 103.4 | 113.1 | 85.0 | 664.3 |

# Supplementary Section 2: Methodological details

## Supplementary Section 2.1: Converged settings for active learning

In the VASP[8–10] simulations, the precision mode was set to "accurate", the Gaussian smearing width was set to 0.05 eV, the energy difference for the break condition of SCF cycles was set to $10^{-8}$ eV and the energy cutoffs defining the plane-wave basis were set to 800 eV for ionic relaxations and to 1050 eV for the active-learning MD in the $NpT$ ensemble. The increase of the cutoff energy by ~30% is meant to avoid Pulay stresses for changing unit cell dimensions arising from thermal expansion as suggested in ref. 11. Convergence of the k-grids and energy cutoffs was tested.

As starting geometries for the active learning training runs we chose the DFT relaxed cells from ref. 6 and optimized the ionic positions for several k-grids (at an energy cutoff of 900 eV) before testing the convergence of the plane wave energy cutoff (ENCUT). The tests showed that 2×3×2 (naphthalene and anthracene) and 2×2×1 (tetracene and pentacene) k-grids and an energy cutoff of 800 eV provided an energy difference of roughly 0.5 meV/atom compared to the highest tested cutoff energy of 1100 eV, as shown in Supplementary Figure 2. The criterion of energy convergence of 0.5 meV/atom applied in ref. 6 to ensure accurate phonon properties from DFT calculations was also chosen here for the purpose of generating the DFT training data.

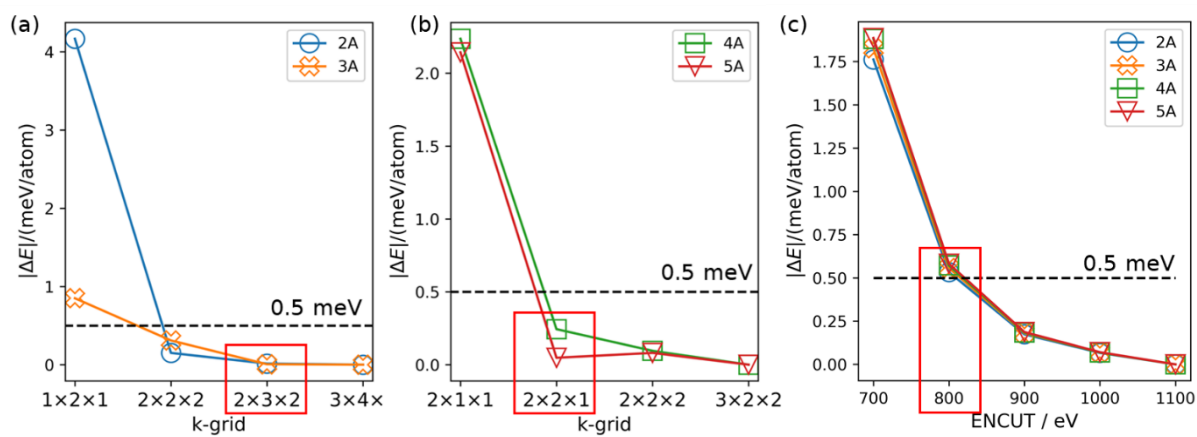

*Supplementary Figure 2: Energy convergence with regards to the k-grid for naphthalene (2A) and anthracene (3A) (a) grids and for tetracene (4A) and pentacene (5A) (b) and convergence with regards to the plane wave energy cutoff (ENCUT) in panel (c) for all systems.*

The number of atoms and the supercell dimensions used in the active learning runs are listed in Supplementary Table 3. Further, the unit cell k-grids from above are adjusted for simulations of super cells by dividing the number of k-points in a direction by the respective super cell dimension and rounding that up to nearest integer. For example, from a 2×3×2 k-grid for the unit cell we get a 2×2×2 grid for a super cell with dimension (1,2,1). Also, the numbers of acquired training configurations and the final maximum force thresholds of the kernel regression are shown.

*Supplementary Table 3: Details of active-learning simulations regarding number of atoms, super cell dimensions, k-grid, collected ab-initio configurations, and reached maximum force threshold at the end of the simulation.*

| acene | $N_{atoms}$ | super cell dimensions | adjusted k-grid | $N$ configs | threshold / (ev/Å) |
|-------|-------------|-----------------------|-----------------|-------------|--------------------|
| 2A | 72 | (1,2,1) | 2×2×2 | 459 | 0.014 |
| 3A | 96 | (1,2,1) | 2×2×2 | 539 | 0.015 |
| 4A | 120 | (2,1,1) | 1×2×1 | 468 | 0.014 |
| 5A | 144 | (2,1,1) | 1×2×1 | 395 | 0.016 |

## Supplementary Section 2.2: Validating Moment Tensor Potentials

## Supplementary Section 2.2.1: Validation data and comparison of energy, forces, and stress

In analogy to the generation of training data, validation data sets were created for each acene in active-learning simulations in an *NpT* ensemble. A temperature gradient between 100 K and 150 K was applied. However, simulations were aborted after 200 ab-initio configurations had been collected. Using energies and forces of these ab-initio configurations, we calculated validation errors for each independently parametrized MTP. The results are listed in Supplementary Table 4 and a direct comparison between DFT and MTP values (including stresses) can be seen in Supplementary Figure 3 for the MTP with the lowest errors in Supplementary Table 4. Energies, forces, and stresses (not included in the table) of these DFT validation configurations sampled between 100 K and 150 K are reproduced exceptionally well by the MTPs, as can further be seen by the Pearson correlation coefficients ($R^2$) plotted in the panels in Supplementary Figure 3. They are almost 1.0 for all properties in all acenes.

*Supplementary Table 4: Validation errors for the three independently parametrized MTPs of each acene. The validation data sets were generated in separated active learning runs and independent of the training sets. The reported errors are the root-mean-square deviations between MTP and DFT energies and forces of the 200 configurations in the validation sets. The MTPs with the lowest errors are highlighted in bold text and those with the smallest force errors were used for the comparisons in Supplementary Figure 3.*

| MTP | energy RMSD / meV | | | force RMSD / [meV/Å] | | |
|-----|------|------|------|-------|------|------|
|     | #1   | #2   | #3   | #1    | #2   | #3   |
| 2A  | 2.59 | **1.65** | 3.01 | 5.67  | **4.93** | 6.60 |
| 3A  | 4.67 | 2.70 | **2.46** | 12.31 | 9.23 | **8.39** |
| 4A  | 5.76 | **4.67** | 5.88 | 11.68 | **8.83** | 12.1 |
| 5A  | 3.09 | 3.00 | **2.70** | **7.69** | 7.73 | 7.96 |

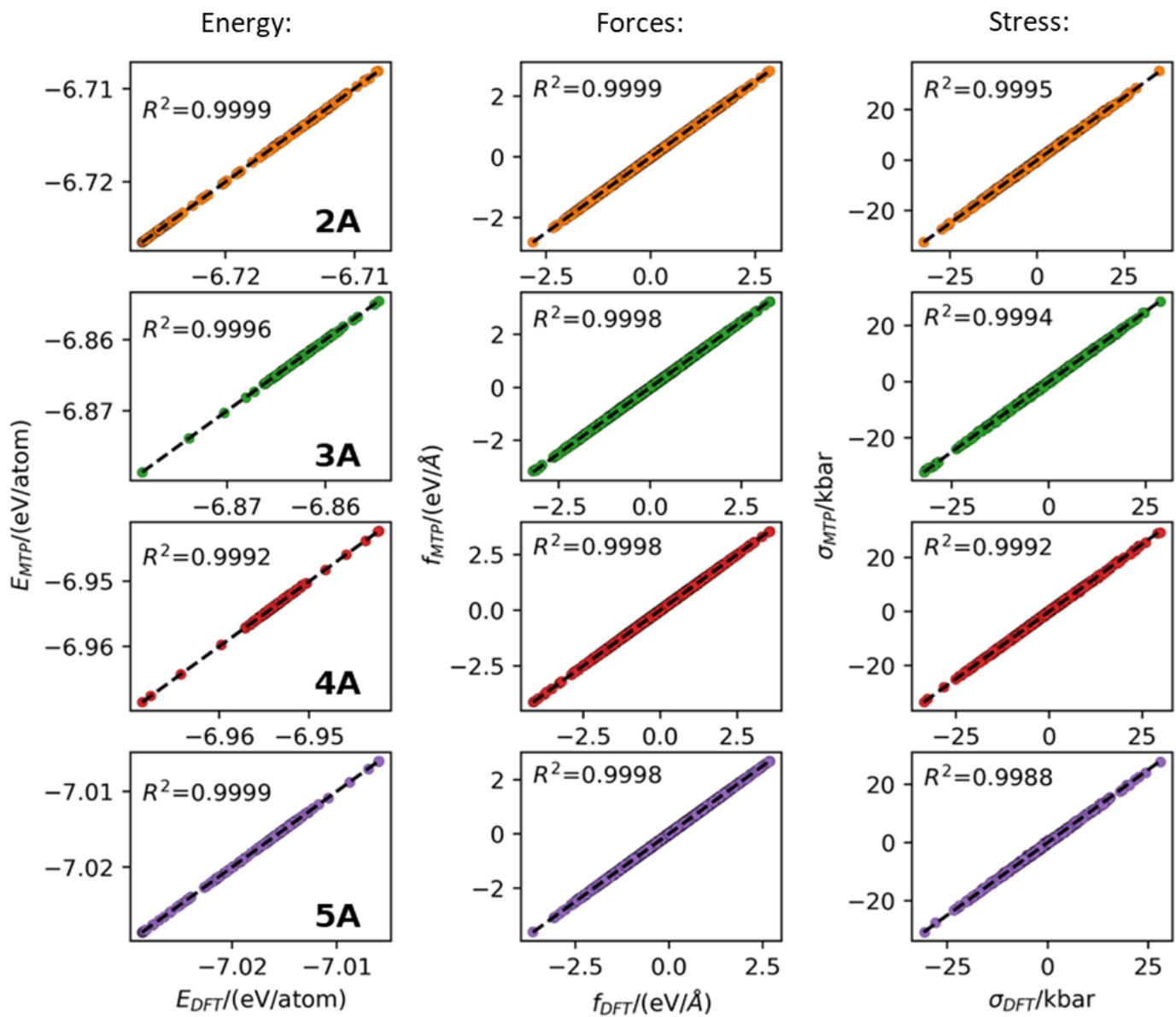

*Supplementary Figure 3: Validation of MTPs. Comparing DFT (x-axes) and MTP (y-axes) calculated energies, forces, and stresses for the validation structures of the studied acenes. The $R^2$ values in the panels are the Pearson correlation coefficients.*

## Supplementary Section 2.2.2: Validating MTPs for describing phonon properties

In the main manuscript, the phonon band structures calculated with DFT and with the MTPs are compared for naphthalene and pentacene. For the sake of completeness, in Supplementary Figure 4 also the analogous plots for anthracene and tetracene are presented. In Supplementary Table 5, the frequency root mean square deviation (RMSD) between the DFT band structures from Kamencek & Zojer[6,12] and those of the three independently parametrized MTPs are listed for each of the materials. Supplementary Figure 5 contains a comparison of the phonon DOSs in the full frequency range.

*Supplementary Table 5: Frequency root mean square deviations (RMSD) between DFT and MTP calculated phonon band structures for each acene. Number in brackets refer to the deviations for bands up to only 5 THz. The frequency RMSDs are shown for each of the three independently parametrized MTPs per system. The MTPs eventually used to perform the Wigner Transport Equation simulations is highlighted by bold numbers. They are equivalent to the MTPs yielding the lowest force errors for the validation structures highlighted in the previous table. For further clarity, we included the same table but present the frequency RMSDs in units of $cm^{-1}$ instead of THz.*

frequency RMSD(DFT,MTP) / THz

| MTP | #1 | #2 | #3 |
|-----|-----|-----|-----|
| 2A | 0.089 (0.068) | **0.073 (0.054)** | 0.11 (0.093) |
| 3A | 0.132 (0.087) | 0.091 (0.040) | **0.082 (0.034)** |
| 4A | 0.107 (0.073) | **0.085 (0.034)** | 0.112 (0.068) |
| 5A | **0.079 (0.045)** | 0.083 (0.049) | 0.086 (0.045) |

frequency RMSD(DFT,MTP) / $cm^{-1}$

| MTP | #1 | #2 | #3 |
|-----|-----|-----|-----|
| 2A | 3.0 (2.3) | **2.4 (1.8)** | 3.7 (3.1) |
| 3A | 4.4 (2.9) | 3.0 (1.3) | **2.7 (1.1)** |
| 4A | 3.6 (2.4) | **2.8 (1.1)** | 3.7 (2.3) |
| 5A | **2.6 (1.5)** | 2.8 (1.6) | 2.9 (1.5) |

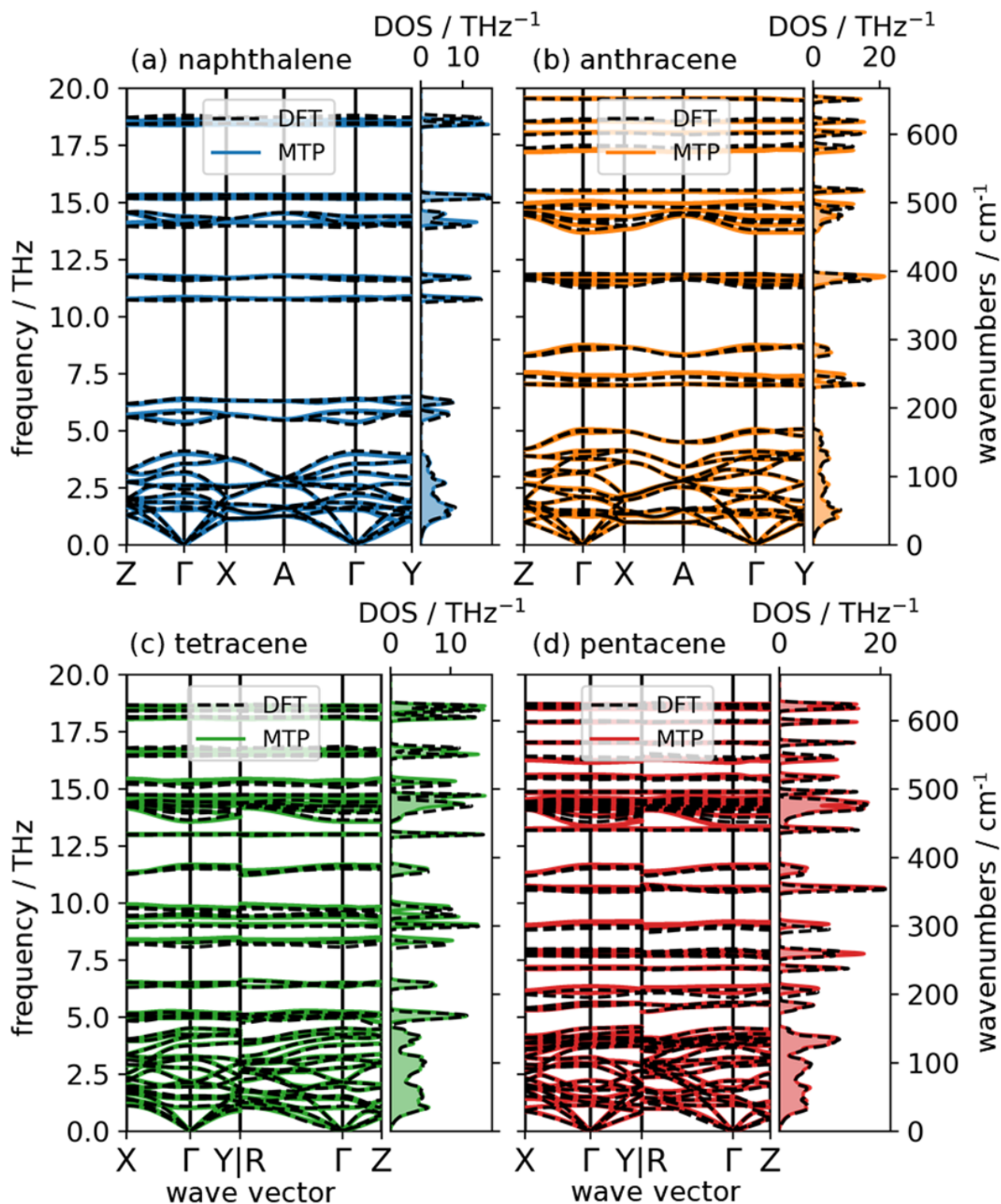

*Supplementary Figure 4: Phonon band structure and density of states of naphthalene(a), anthracene (b), tetracene (c), and pentacene (d) calculated with PBE+D3BJ (black, dashed line) from ref. 6 and with a system-specifically trained MTP (colored, solid lines). The frequency axes are provided in units of terahertz and wavenumbers. In passing we note that the lengths of the high symmetry paths are not identical for different materials. Data from ref. 6 have been published under CC-BY 3.0 license by the Royal Society of Chemistry; copyright, the authors.*

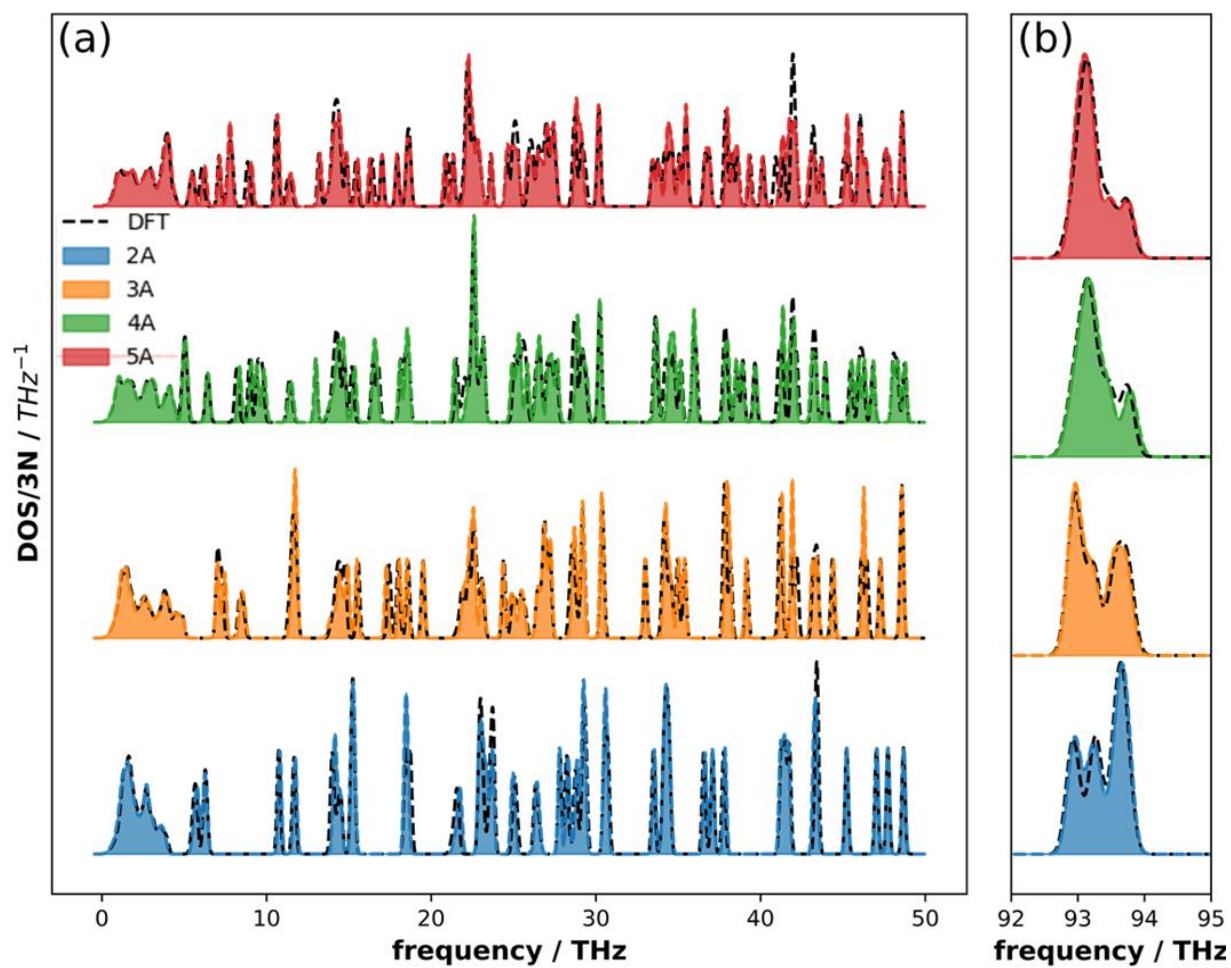

*Supplementary Figure 5: Phonon density of states (DOS) of naphthalene (blue), anthracene (orange), tetracene (green), and pentacene (red) calculated with the best MTPs. For the sake of comparison, the dashed black lines show the respective DFT-calculated DOSs. For plotting the DOS a broadening of 0.1 THz was applied.*

*Supplementary Table 6: The reduced coordinates of the high-symmetry points in the respective first Brillouin zones of naphthalene (2A), anthracene (3A), tetracene (4A), and pentacene (5A) used in SI-Figure 4 and Figure 2 are shown for the representative labels ("X", "A", etc.).*

| label | 2A | 3A | 4A | 5A |
|-------|----|----|----|----|
| X | (-1/2,0,0) | (-1/2,0,0) | (1/2,0,0) | (1/2,0,0) |
| Y | (0,0,1/2) | (0,0,1/2) | (0,1/2,0) | (0,1/2,0) |
| Z | (0,1/2,0) | (0,1/2,0) | (0,0,1/2) | (0,0,1/2) |
| A | (-1/2,1/2,0) | (-1/2,1/2,0) | - | - |
| R | - | - | (-1/2,1/2,1/2) | (-1/2,1/2,1/2) |

The lattice parameters of the unit cells for which the phonons were calculated are presented in Supplementary Table 2 and the reduced coordinates of the high-symmetry points for the band structures presented in Supplementary Figure 4 and Figure 2 of the main work are shown in Supplementary Table 6.

In Supplementary Figure 6 the eigenvector overlap between DFT- and MTP-calculated phonon modes is shown for each studied acene. The MTP calculations have been performed for the potentials with the smallest frequency RMSD (see Supplementary Table 5). Most modes are reproduced well and their overlap lies above 95%. For the two triclinic systems 4A and 5A there is a significant number of modes between 50% and 100%. The histograms on the right side of the panels further show the relative number of modes for an overlap of at least 95% (highest bar). These histograms are normalized to the number of phonon modes (108, 144, 180, and 216). The overlap should further show that the MLP applied here can reproduce DFT to a satisfactory degree. Details on how this eigenvector overlap is determined can be found in the Supporting Information, Section 12 of ref. 13 and in ref. 14.

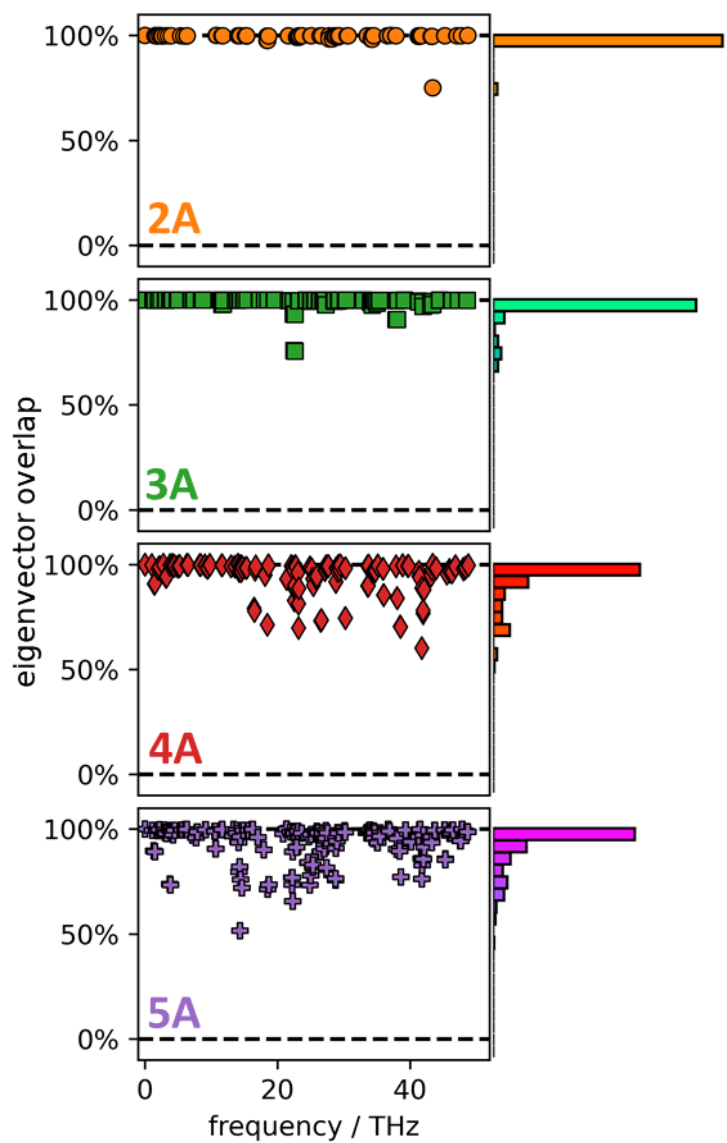

*Supplementary Figure 6: Eigenvector overlap between DFT- and MTP-calculated phonon modes as a function of vibrational frequencies at the Γ-point for 2A, 3A, 4A, and 5A. On the right side of each panel a histogram shows their distribution of the overlaps.*

## Supplementary Section 2.3: Estimated computational cost of 3$^{rd}$-order force constants from DFT

Here we want to clarify the statement that calculating 3$^{rd}$-order force constants with DFT would be computationally intractable for the systems at hand in the Results section of the main manuscript. What follows is a cost comparison between simply using DFT and creating system-specific MLPs as surrogate models to DFT for the task of anharmonic lattice dynamics instead. For example, calculating the 3$^{rd}$-order force constants of pentacene (746,712 single-point calculations) on the infrastructure available to us Vienna Scientific Cluster, "VSC-5": AMD EPYC 7713 @ 2 GHz) would cost roughly 65 million core-h. In practice this simulation would fully occupy the VSC-5 for 30 days, which is at the time of writing Austria's most powerful computation cluster (benchmark value 2.31 PFlop/s).[15] In contrast, acquiring the training data, parametrizing three individual MTPs and performing test and validation simulations amounted to only 13 thousand core-h. For the smaller acenes the number of required displacements is smaller, however the number of atoms in those super cells is similar to that of pentacene. Therefore, even for naphthalene we estimate a cost of 9 million core-h.

## Supplementary Section 2.4: Converging Wigner Transport Equation results

First, we tested the difference between the "full", linearized BTE[16] and the BTE in the relaxation time approximation (RTA).[17] The difference is rather small, as can be seen in Supplementary Figure 7 for naphthalene, with a 6×7×6 $q$-mesh. Therefore, we conducted the computationally more demanding phono3py simulations of the larger acenes all within the relaxation time approximation.

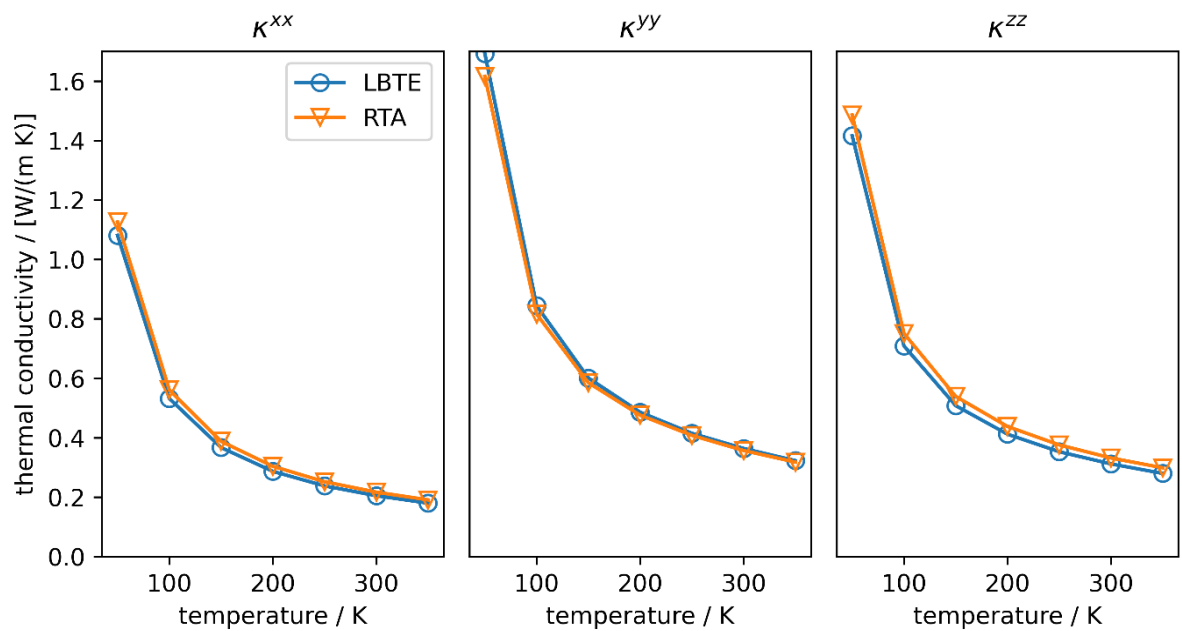

*Supplementary Figure 7: Thermal conductivity tensor elements $\kappa^{xx}$, $\kappa^{yy}$, and $\kappa^{zz}$ of naphthalene over temperature calculated with a direct solution of the linearized Boltzmann transport equation[16] ("LBTE"; blue circle) and within the relaxation time approximation[17] ("RTA"; orange triangles).*

Secondly, we tested the impact of the displacement amplitude for the 3$^{rd}$-order force constant calculation by comparing temperature-averaged, isotropic thermal conductivities for the case of naphthalene. We choose to compare the mean thermal conductivity between 50 K and 500 K because the variation of the thermal conductivity with displacement amplitude has a higher magnitude towards lower temperatures. This way, the value we compare is more reflective of the entire temperature range and not arbitrarily fixated a on specific temperature (e.g. 300 K). The results show that convergence is reached at a value of 0.03 Å (see Supplementary Figure 8), which is also the default value of phono3py. This value was then chosen for all further calculations.

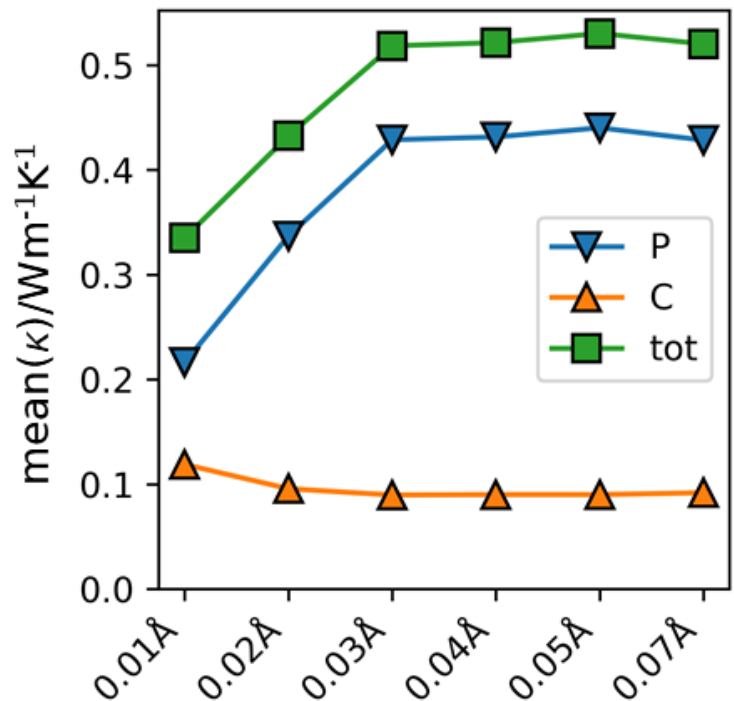

*Supplementary Figure 8: Temperature-average, isotropic thermal conductivity of naphthalene for a range of displacement amplitudes. Upside triangles refer to the coherence contribution, downside triangles to populations' contribution, and squares to their sum, the total thermal conductivity.*

Next, we converged the thermal conductivity with regards to the super cell dimensions for the 3$^{rd}$-order force constant calculations. While varying the super cell dimensions for calculating the 3$^{rd}$-order force constants, the dimensions of the also required harmonic force constants stayed the same. Those were calculated for super cells with a converged (2,3,2) dimension (based on ref. 13). The resulting anisotropic thermal conductivities are presented in Supplementary Figure 9 as a function of temperature. Based on this test, we applied the same super cell dimensions that led to converged harmonic vibrations also for calculating 3$^{rd}$ order force constants in all systems. Those dimensions can be seen in Table 2 of the main text.

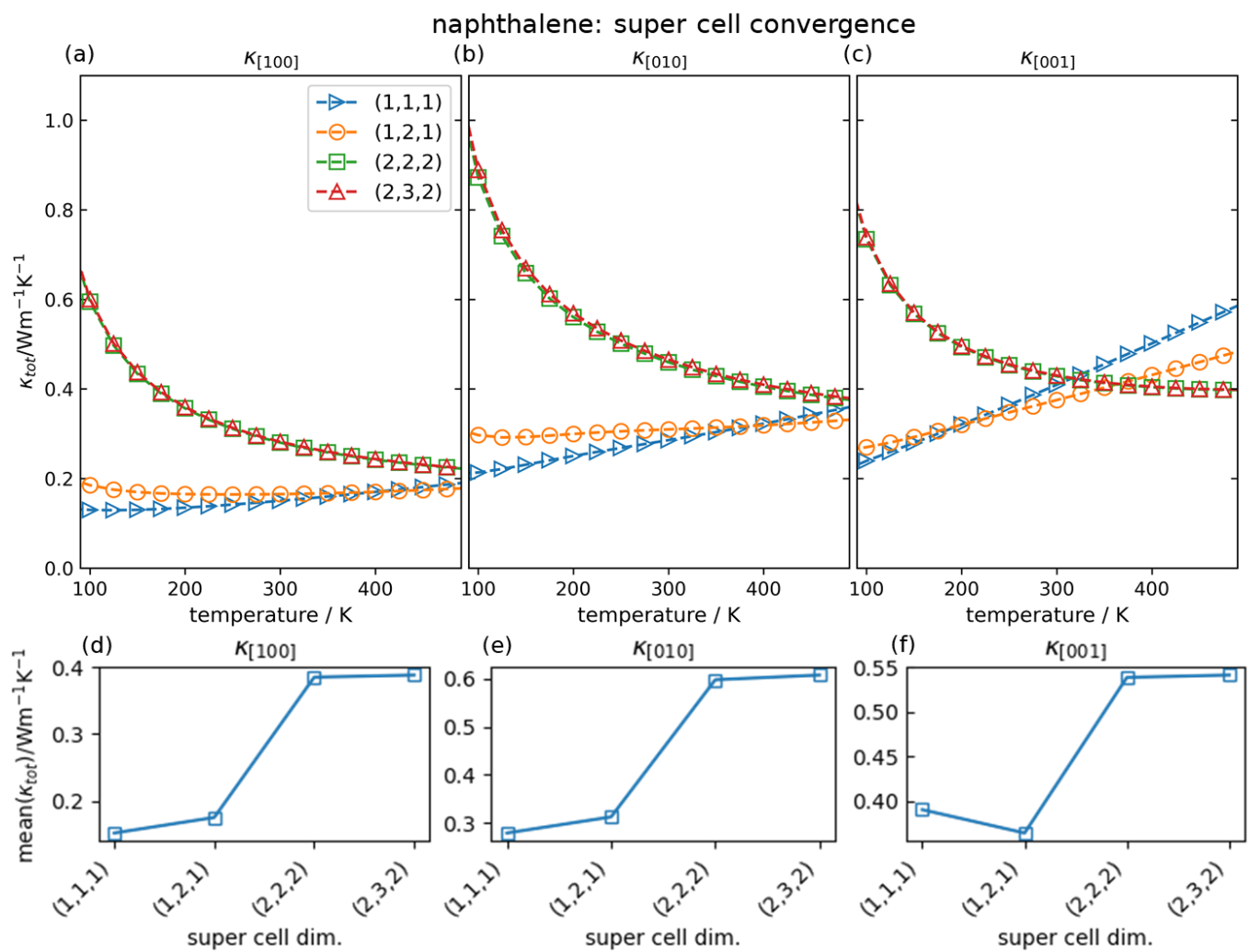

*Supplementary Figure 9: Thermal conductivity of naphthalene in the three crystal directions [100], [010], and [001] as a function of temperature (a-c), where different colors and symbols denote the super cell dimensions used for calculating the 3rd-order force constants. The size of the supercells is specified in the legend in panel (a). In panels (d-f) the temperature-averages of the thermal conductivity as a function of the super cell dimensions are shown.*

Finally, we tested several *q*-point meshes for calculating the three-phonon scattering processes in each of the four acenes individually. The grids were chosen such that they are (close-to) equidistant in the three spatial directions. These tests can be seen for naphthalene in Supplementary Figure 10, anthracene in Supplementary Figure 11, tetracene in Supplementary Figure 12, and pentacene in Supplementary Figure 13. Without discussing the tests in detail, we note that the thermal conductivity converges faster (smaller required *q*-point density) for some directions and that the directions cannot be treated as fully independent, probably due to the triclinic angles.

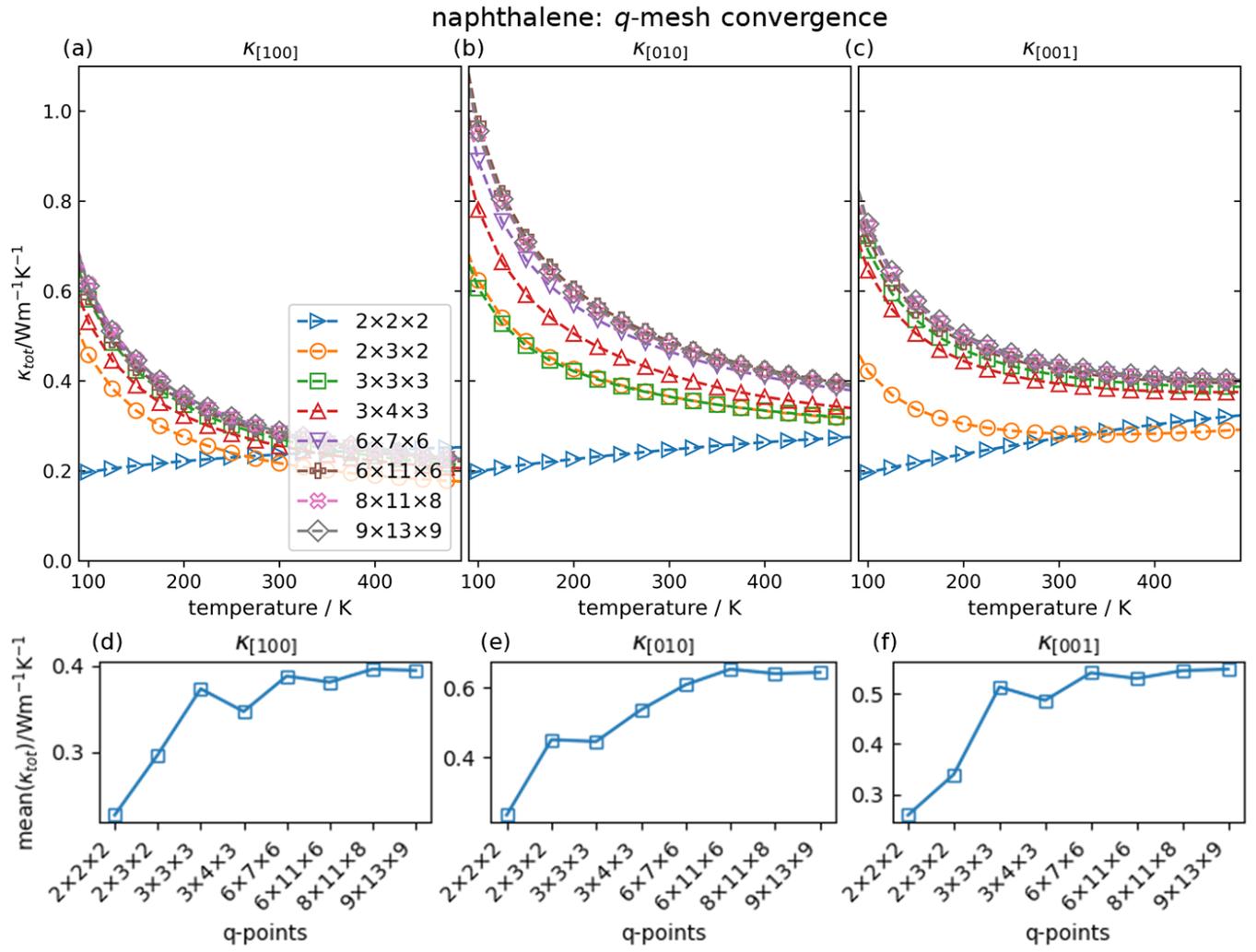

*Supplementary Figure 10: Thermal conductivity of naphthalene in the three crystal directions [100], [010], and [001] as a function of temperature (a-c), where different colors and markers denote the applied q-mesh for solving the WTE. The meshes are specified in the legend in panel (a). In panels (d-f) the temperature-averages of the thermal conductivity as a function of the q-mesh are shown.*

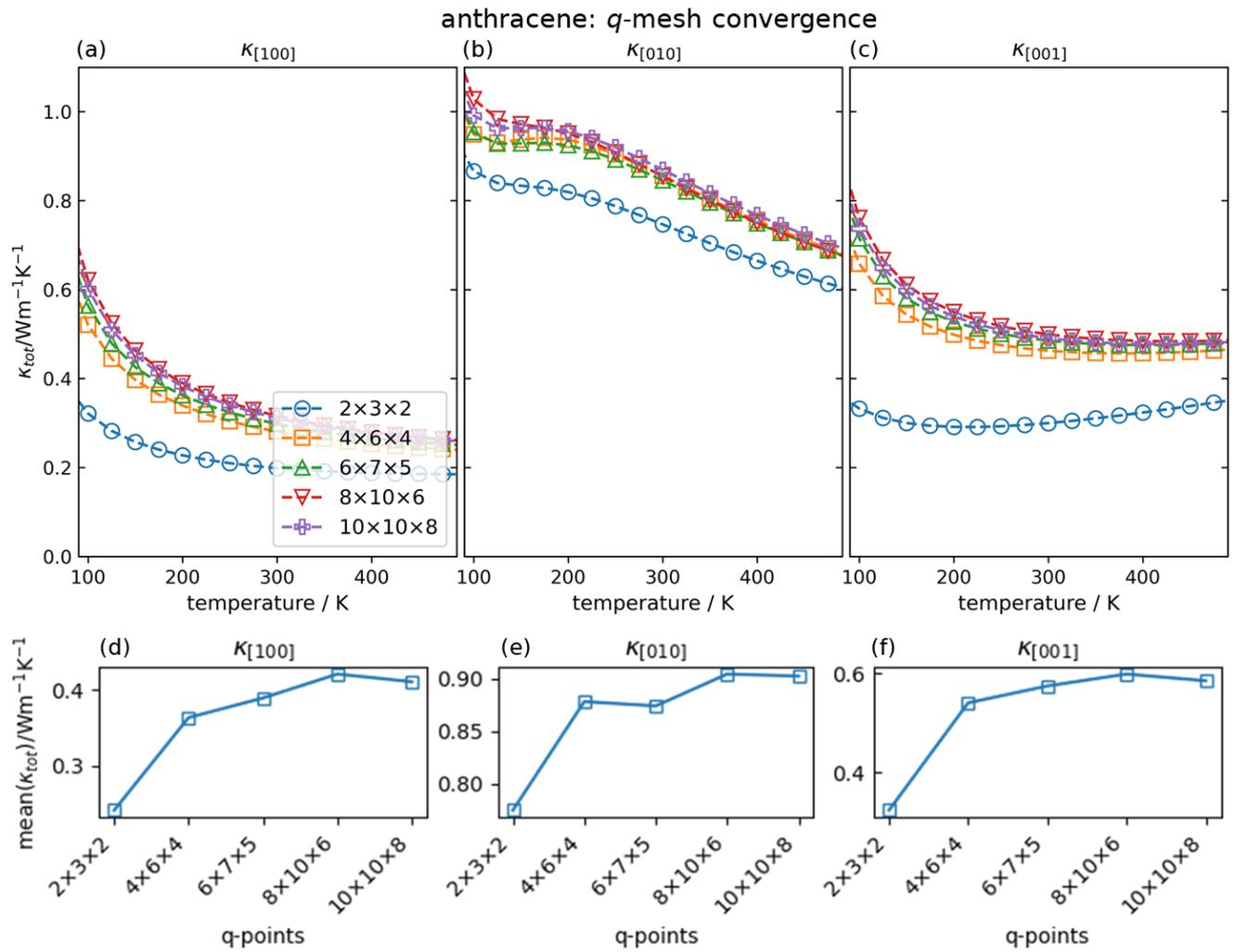

*Supplementary Figure 11: Thermal conductivity of anthracene in the three crystal directions [100], [010], and [001] as a function of temperature (a-c), where different colors and markers denote the applied q-mesh for solving the WTE. The meshes are specified in the legend in panel (a). In panels (d-f) the temperature-averages of the thermal conductivity as a function of the q-mesh are shown.*

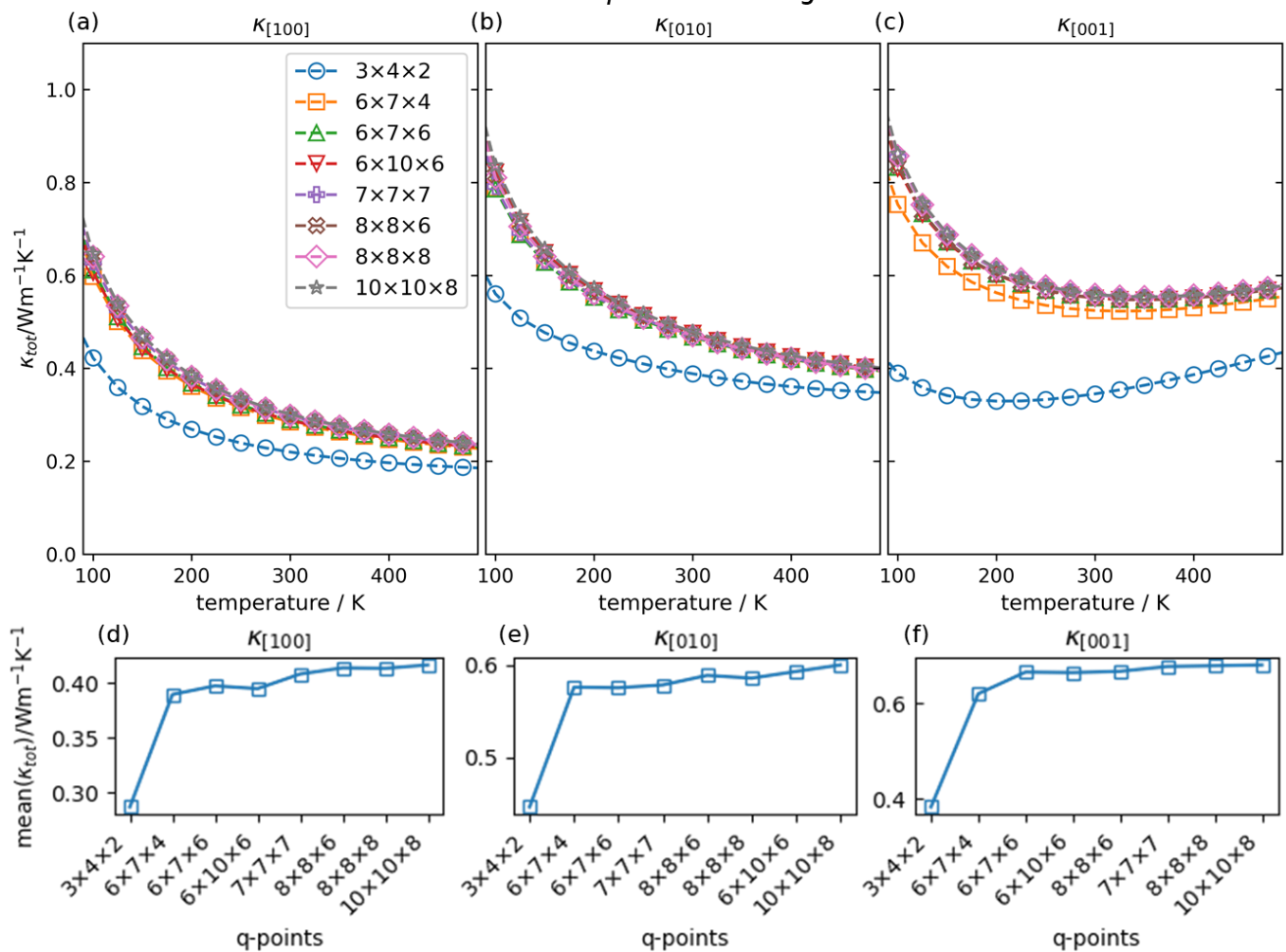

*Supplementary Figure 12: Thermal conductivity of tetracene in the three crystal directions [100], [010], and [001] as a function of temperature (a-c), where different colors and markers denote the applied q-mesh for solving the WTE. The meshes are specified in the legend in panel (a). In panels (d-f) the temperature-averages of the thermal conductivity as a function of the q-mesh are shown.*

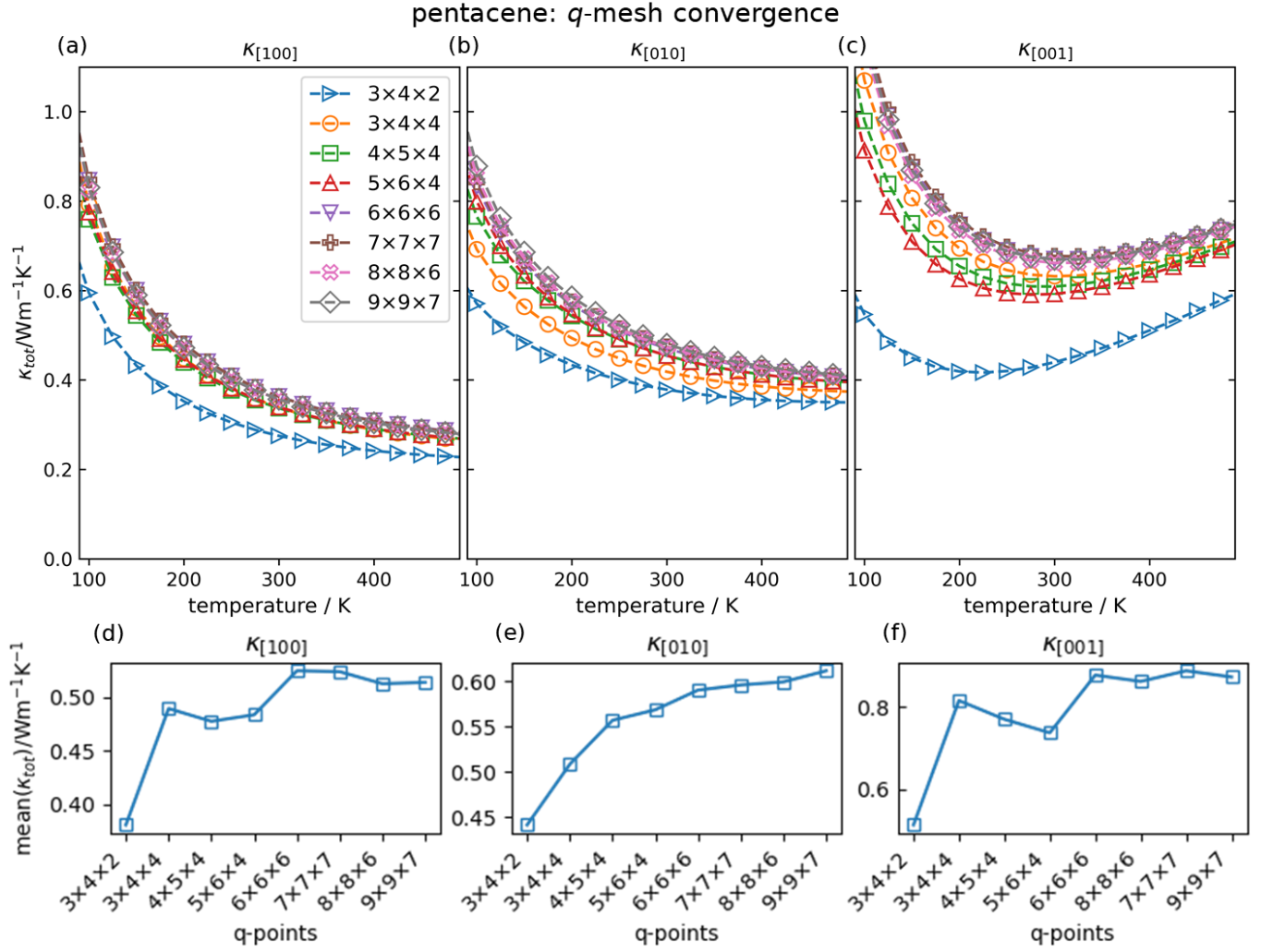

*Supplementary Figure 13: Thermal conductivity of pentacene in the three crystal directions [100], [010], and [001] as a function of temperature (a-c), where different colors and markers denote the applied q-mesh for solving the WTE. The meshes are specified in the legend in panel (a). In panels (d-f) the temperature-averages of the thermal conductivity as a function of the q-mesh are shown.*

Based on these tests we chose the *q*-meshes for the results presented in the main text. There they are listed in Table 2.

## Supplementary Section 2.5: Phonon lifetimes of the acenes

Phonon modes with lifetimes smaller than the inverse of their angular frequencies $\omega$ are considered to be overdamped according to the Ioffe-Regel limit in time.[18] In that case, the quasi-particle picture of phonons breaks down and one should not apply the WTE or BTE for that matter. To ensure that the phonon modes in the acenes are not overdamped we looked at their lifetimes as a function of their vibrational frequency (Supplementary Figure 14) and found that the lifetimes of all phonons relevant for heat transport are above $1/\omega$. The C-H vibrations found around 95 THz (~3170 cm⁻¹) are omitted.

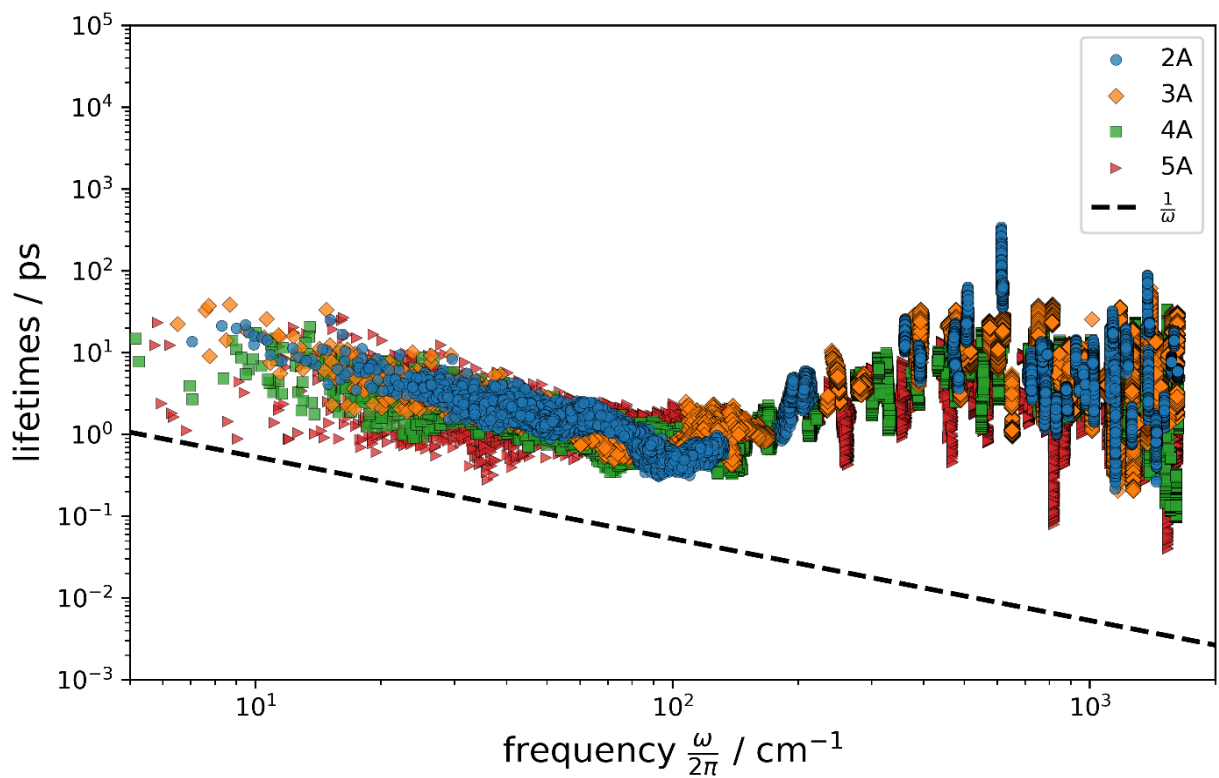

*Supplementary Figure 14: Phonon lifetimes at 300 K as a function of vibrational frequency of naphthalene (circles), anthracene (diamonds), tetracene (squares), and pentacene (triangles). The dashed line represents a lifetime of $\tau = 1/\omega$ below which modes would be overdamped, according to the Ioffe-Regel limit in time.[18]*

## Supplementary Section 2.6: The unusual temperature evolution of the thermal conductivity of anthracene in the [010] direction

In Figure 6b of the main work $\kappa_P(T)$ of 3A along [010] has not been shown, as it does not follow the trend of the other acenes and displays an unconventional bump around 200 K. In this chapter, we included it in Supplementary Figure 15b and discuss several methodical attempts to impact this temperature evolution.

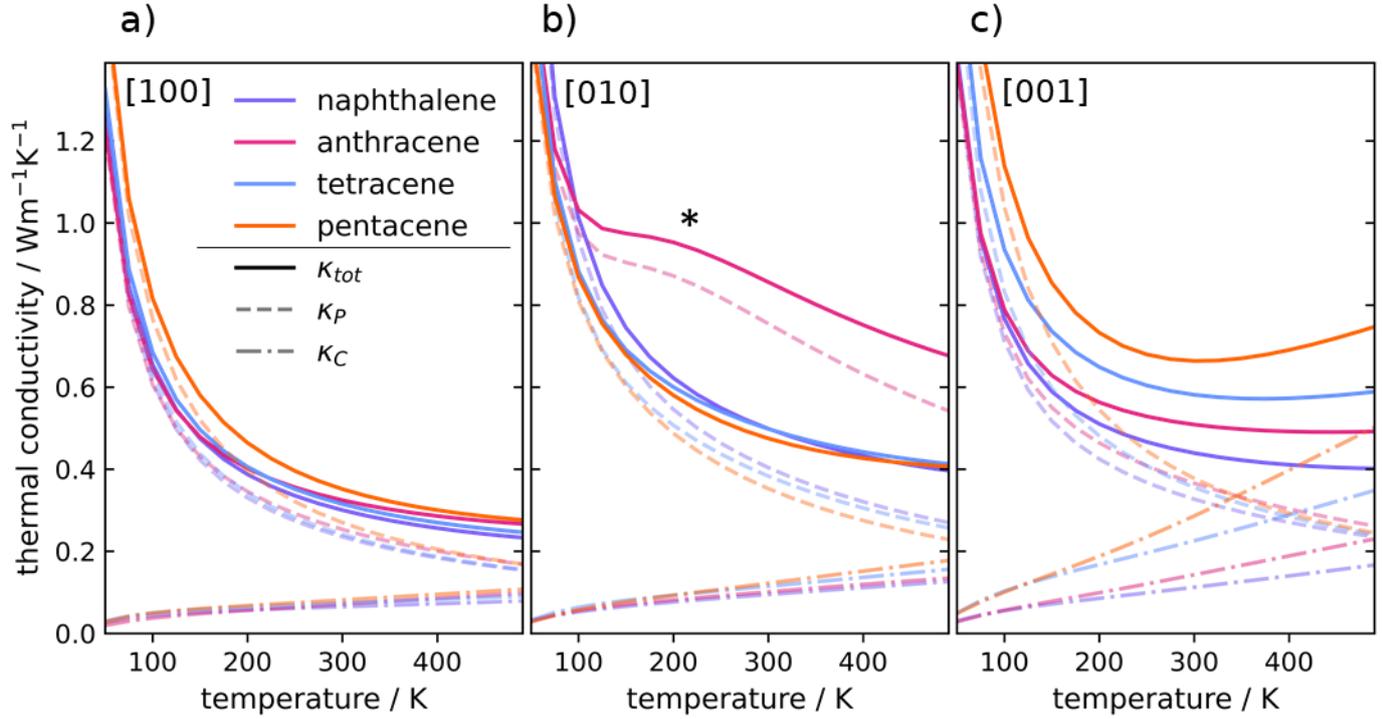

*Supplementary Figure 15: Anisotropic lattice thermal conductivities of naphthalene (purple), anthracene (magenta), tetracene (blue), and pentacene (orange) as a function of temperature. Solid lines show the total thermal conductivity $\kappa_{tot}$, dashed lines the propagation contribution $\kappa_P$ and dash-dotted lines the tunneling contribution $\kappa_C$. The anisotropic conductivities according to eq.(3) are presented for the crystal directions [100] in panel a), for [010] in panel b), and for [001] in panel c).*

Several parameters in the course of the procedure followed for calculating the lattice thermal conductivity with phono3py have been tested, but to no avail. By inspecting the cumulative $\kappa_P^{[010]}$ (Supplementary Figure 16) we can say that the culprits are supposedly overestimated lifetimes of optical phonon modes above 10 THz that only affect $\kappa_P^{[010]}$ due to the peculiarities of the phonon dispersion (see Supplementary Figure 4b; Z-Γ path compared to Γ-X or Γ-Y around 14 THz).

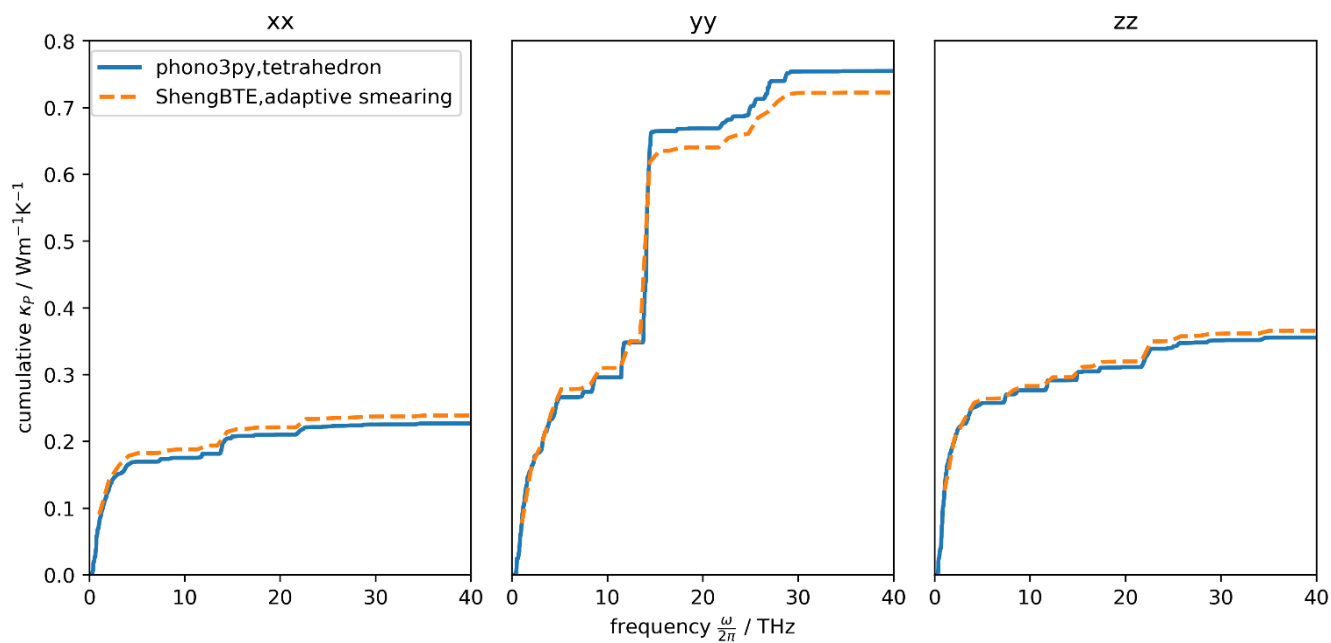

*Supplementary Figure 16: Comparison of the cumulative propagation thermal conductivities of anthracene at 300 K as a function of vibrational frequency obtained with the tetrahedron method (blue line) from phono3py and an adaptive Gaussian smearing approach (dashed, orange line) from ShengBTE.[19] The three panels from left to right present the diagonal elements "xx", "yy", and "zz" of the thermal conductivity tensor.*

For the [010] crystal direction, the choice of the Brillouin zone integration method strongly impacts $\kappa_P$ as can be seen in Supplementary Figure 17. This is in sharp contrast to the observations for other directions (Supplementary Figure 17d and f) and to all directions in naphthalene (Supplementary Figure 17a-c). In detail, phono3py allows the use of different mathematical algorithms for performing the Brillouin-zone integration: for the tetrahedron method the integration is performed numerically within small volumes of the irreducible part of the Brillouin zone, while for the Gaussian smearing method one replaces the Dirac delta function with a Gaussian for the purpose of smooth integration. The latter requires to choose a value for the standard deviation of the Gaussian function. For the smallest standard deviation $\sigma$ of 0.05 THz, thermal conductivities are almost identical to the tetrahedron method. The largest two tested smearing parameters of 1.0 THz and 1.5 THz yield very similar thermal conductivities. Due to this agreement between $\sigma$=1.0 THz and $\sigma$=1.5 THz and the exploding computational costs for calculating the ever-larger number of phonon interactions that arise with increasing $\sigma$, 1.5 THz was the largest chosen value. With this $\sigma$, $\kappa_P^{[010]}$ drops monotonic and at 300 K it is roughly 40% smaller than for the tetrahedron method. To check whether the integration methods would also produce diverging results, the same test was performed for naphthalene. As indicated already above, the Gaussian smearing and tetrahedron methods produce largely identical results for naphthalene (Supplementary Figure 17a-c). Only for [001] (panel c) a minor difference between tetrahedron (blue dots) and $\sigma$=1 THz (purple circles) can be seen.

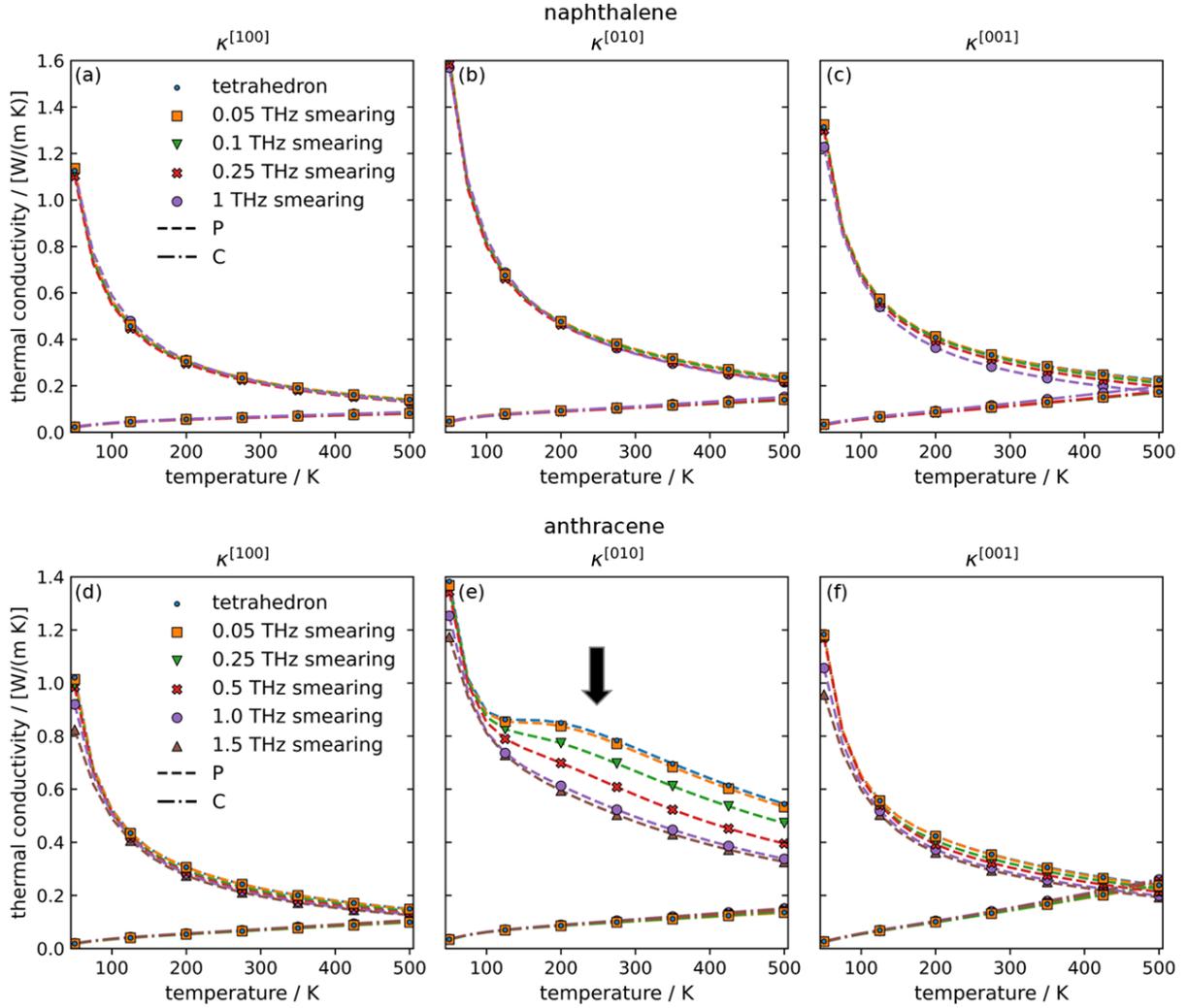

*Supplementary Figure 17: Anisotropic thermal conductivities of naphthalene (a-c) and anthracene (d-f) in the crystal directions [100] (a,d), [010] (b,e), and [001] (c,f). Dashed lines refer to $\kappa_P$ and dash-dot lines to $\kappa_C$. Symbols and colors are used for data from either the tetrahedron method (blue dots) or Gaussian smearing with varying $\sigma$-parameter given in THz (see legends in panels a and d) in the Brillouin-zone integration. The employed q-meshes for naphthalene and anthracene for these tests were 6×7×6 and 6×7×5, respectively and our tests in section 2.4. show that especially for the tetrahedron method, those meshes are close to converged. The cutoff for the Gaussian smearing calculations was set to 2.5 standard deviations. The $2^{nd}$- and $3^{rd}$-order force constants were calculated for converged (2,3,2) super-cells and atomic displacement of 0.03 Å for both materials.*

We also tested whether using a different software all together would yield a different outcome by repeating the simulations with the ShengBTE code.[19] For that purpose, the phono3py[17,20] force constants were transformed into the ShengBTE format using the hiphive package.[21] Additionally, the default integration method of ShengBTE (adaptive Gaussian smearing)[19] was chosen for this simulation. The resulting propagation thermal conductivity values ($\kappa_P$) vary only slightly from those obtained with phono3py and the tetrahedron method, using a 6×7×5 $q$-mesh for both simulations. For the diagonal elements of the tensor we get 0.239 (0.229) Wm$^{-1}$K$^{-1}$, 0.723 (0.743) Wm$^{-1}$K$^{-1}$, and 0.366 (0.358) Wm$^{-1}$K$^{-1}$at 300 K with ShengBTE (phono3py). In Supplementary Figure 16 the cumulative thermal conductivity is plotted as a function of vibrational frequency for the three diagonal elements of the tensor. Most noteworthy; the functions are almost identical and $\kappa^{yy}$ exhibits a major jump around 14 THz in both methods. This sudden jump hints that the corresponding phonon lifetimes are rather high and could be overestimated.

# Supplementary Section 3: Temperature dependent $\Gamma$-point linewidths

The anharmonic phonon linewidths calculated with phono3py in the relaxation time approximation are an integral ingredient (via the directly related phonon lifetimes) for calculating the thermal conductivities of the acenes. Therefore, to estimate the validity of those linewidths, we compared the temperature induced broadening of linewidths at the $\Gamma$-point to several experiments on these materials. Contrary to many sections in the main text, here we are discussing vibrations in the unit of cm$^{-1}$ instead of THz, as this is the unit predominantly used in spectroscopy. In Supplementary Figure 18 the experimental data are plotted as open circles and the theoretical linewidths as lines. However, the comparison is not straightforward since the calculated phonon modes need to be matched to those presented in experimental works. Modes for anthracene and pentacene were matched based on the irreducible representation provided in with the experiments (supporting information of Asher et al.)[22] and based on the similarity of frequencies. This invites some ambiguity for pentacene with only one representation ("$A_g$") and close-lying intermolecular phonon bands, but for naphthalene and anthracene the additional "$B_g$" presentation simplified the identification. Experiments on naphthalene have been taken from Bellows & Prasad[23] and Hess & Prasad,[24] where the irreducible representation was not specified. Here, for the experimental modes at 121 and 141 cm$^{-1}$ only one mode was eligible in both cases and based on the tendency of the theoretical frequencies being slightly lower, the lower two modes were matched, disregarding theoretical modes $A_g^2$ and $B_g^2$ at 82 and 87 cm$^{-1}$. Furthermore, in experiments there are additionally frequency shifts due to thermal expansion and potentially phonon renormalization which are completely neglected in the simulations and outside the scope of this work. Anharmonic linewidths from phono3py match the experimentally determined ones to a certain degree. For naphthalene the linewidth of the 141 cm⁻¹ mode is underestimated and that of the 764 cm⁻¹ mode displays non-trivial behavior as two modes combine due to frequency shifts, that cannot be fully captured by theory. For anthracene all linewidths are underestimated, regardless of whether the tetrahedron (dash dotted blue line) or Gaussian smearing method (dashed orange line) applied. Also, it appears that the experimental linewidths increase stronger than linear after 150 K which could hint at higher-order scattering terms becoming relevant which could explain the (presumably) overestimated lifetimes of some modes we faced for that material (see Supplementary Section 2.6). For pentacene, for the two higher-lying modes, one again observes larger widths in the experiments than in the simulations. Only the two lowest-frequency modes considered here display a different trend.

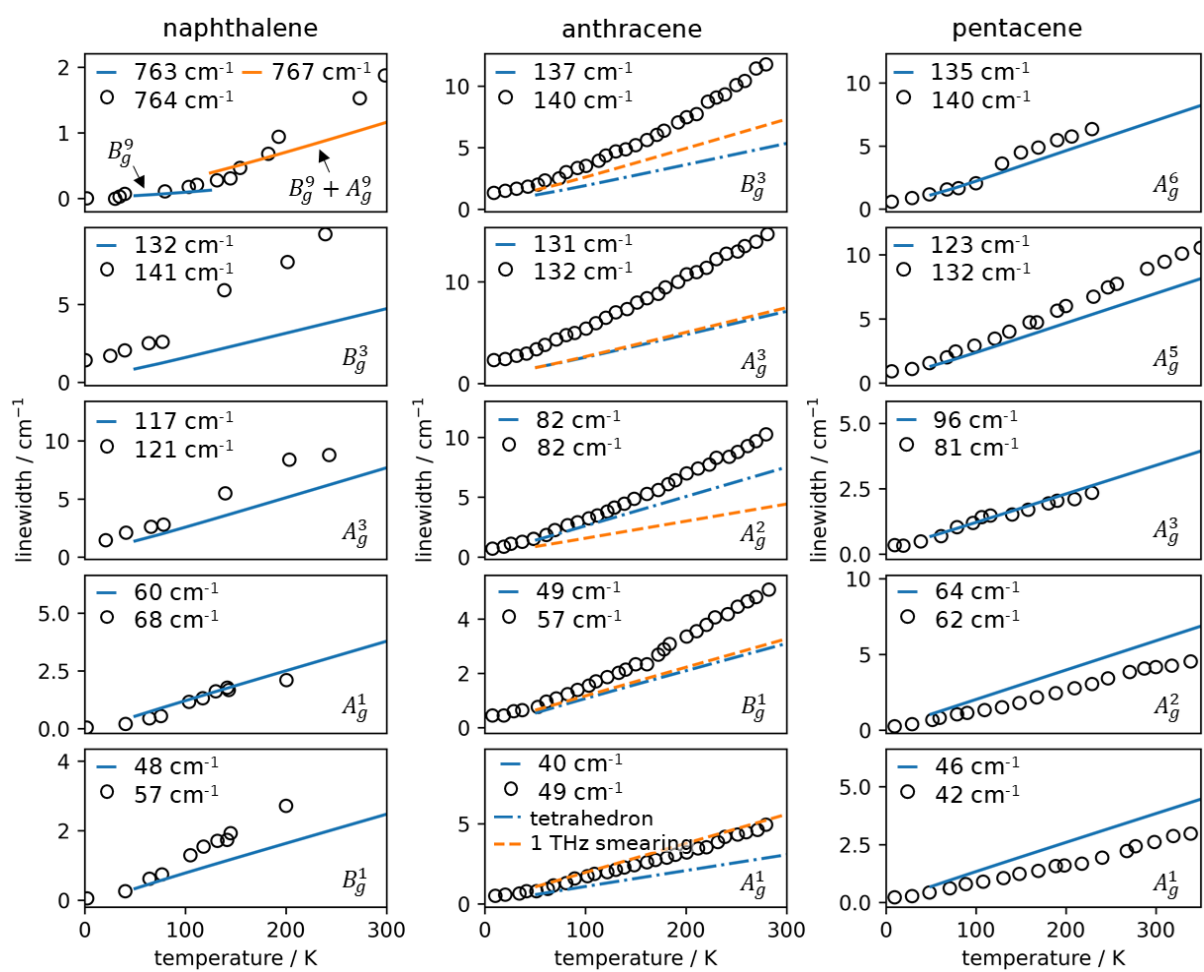

*Supplementary Figure 18: Anharmonic phonon linewidths of naphthalene, anthracene, and pentacene as a function of temperature. The open circles refer to experimentally determined linewidths and the mode frequencies in cm⁻¹ are printed in the respective panel. For naphthalene the experiments are from Bellows & Prasad[23] for the low-frequency modes up to 141 cm⁻¹ and Hess and Prasad[24] for the modes at 764 cm⁻¹. For anthracene and pentacene, the experimental linewidths were published by Asher et al. in the Supporting Information of ref. 22. The colored lines are the respective linewidths from phono3py at the Γ-point. For all three materials results obtained with tetrahedron method for reciprocal-space integration and for anthracene additionally of results with a Gaussian smearing of*

*σ=1 THz are presented. The legends in the panel also specify the measured and calculated frequencies of the modes. Labels of the irreducible representations of the respective modes can be seen in the bottom right corner of each panel.*

Generally, a larger width in the experiments is not entirely unexpected, as there in addition to homogeneous, line broadenings from phonon-phonon scattering also inhomogeneous broadenings occur. These can, for example be triggered by (static) disorder and defects, which increase with temperature. Besides that, also some impact of higher than 3rd order contributions to the anharmonicities could potentially cause larger linewidths in the experiment than in the simulation.

# Supplementary Section 4: Isotropic lattice thermal conductivity of naphthalene

The temperature range in Figure 3a of the main work is adjusted to the range of the available experimental data. A larger temperature range is presented in Supplementary Figure 19.

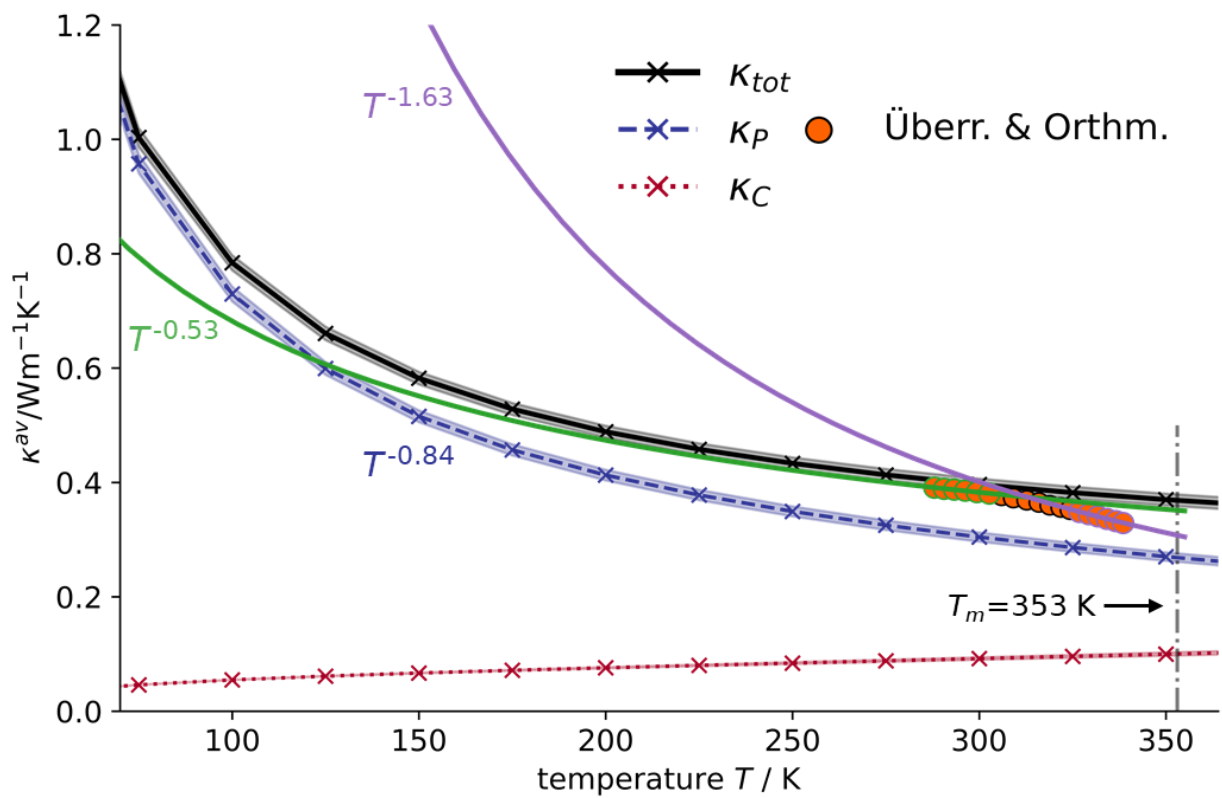

*Supplementary Figure 19: Isotropic lattice thermal conductivity of naphthalene as a function of temperature. Lines and shaded areas refer to the simulation results obtained in this work. The shaded areas show the isotropic thermal conductivities within upper and lower bounds according to eq. (2) in the main work, whereas the lines serve as a guide to the eye and the crosses denote the temperatures for which the WTE has been evaluated. Solid black lines refer to the total thermal conductivity, dashed*

*blue lines denote the propagation contribution $\kappa_p$, and dotted red lines show the additional contributions due to phonon tunneling $\kappa_c$. Experimental results from ref. 25 are shown as orange squares. Two fitted lines for the first (green) and the last (purple) six data points of the experimental thermal conductivity are drawn and exponents of the fits, also for $\kappa_p$, are indicated. The melting temperature of naphthalene at 353 K is represented by a dash-dotted horizontal line.*

## Supplementary Section 4.1: Grain boundary scattering in naphthalene

As the grain-sizes of the naphthalene powder in the experiments of Überreiter and Orthmann[25] is not known, in the following the impact of different arbitrarily chosen grain sizes on the thermal conductivity of naphthalene shall be tested. Overall, the following discussion shows that grain boundary scattering has only a minor influence on the lattice thermal conductivity of naphthalene for mean grain sizes above 10 nm ($10^{-2}$ μm). The results presented in the main work (Figure 3a) are for the default mean grain size of 1 m ($10^6$ μm), i.e., for neglecting grain-boundary scattering effects. In Supplementary Figure 20 the (arithmetic) average thermal conductivity of naphthalene is shown for several mean grain sizes. The blue, orange, green, and red lines correspond to $10^6$, $10^2$, $10^1$, and $10^0$ μm and are hardly distinguishable. The purple line corresponding to $10^{-1}$ μm is only noticeably lower between 50 K and 250 K.

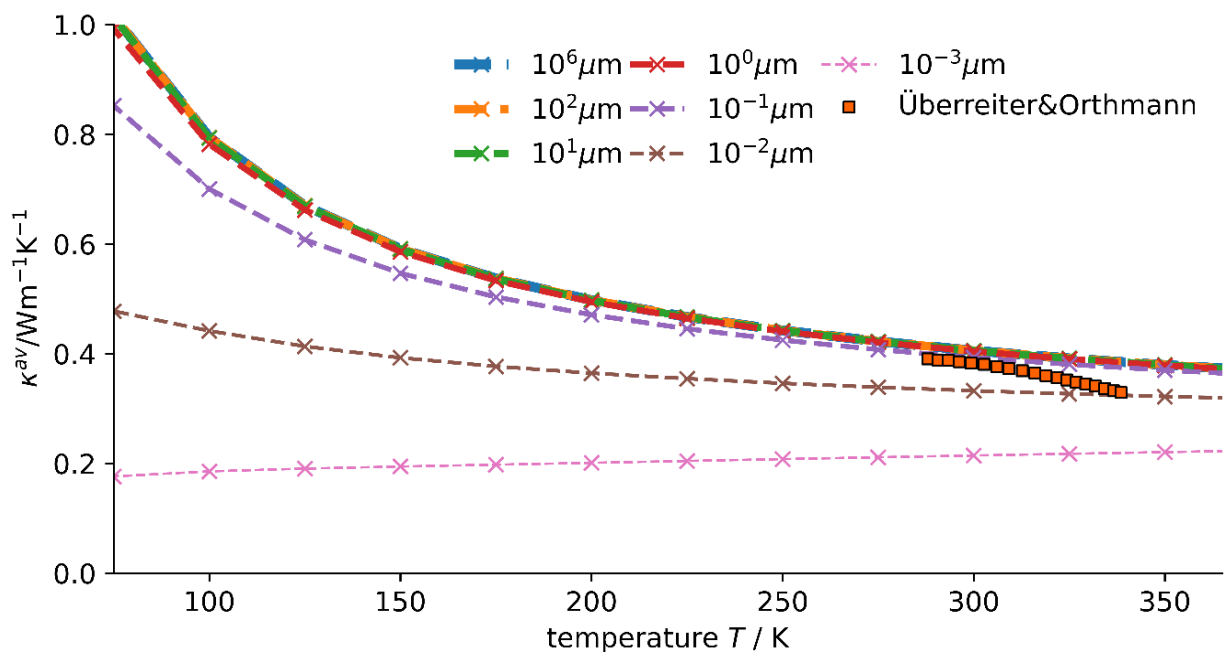

*Supplementary Figure 20: Isotropic lattice thermal conductivity of naphthalene as a function of temperature for various grain boundary scattering settings (i.e. mean grain sizes). The considered grain sizes are listed in the legend. For comparison, the experimental data points of Überreiter and Orthmann[25] are printed as orange squares.*

## Supplementary Section 4.2: Impact of the thermal expansion

We additionally calculated a second set of 2[nd] and 3[rd] order force constants for super cells of the experimental[2] unit cell of naphthalene at 295 K to estimate the impact of thermal expansion on the thermal conductivities of naphthalene. This was to ensure that the observed trends are not affected by the choice of the unit cell, i.e. its volume. The corresponding thermal conductivities were compared to those of the unit cell with DFT-relaxed lattice parameters already presented in the main work. These lattice parameters are close to those experimentally determined by single crystal neutron diffraction at 5 K.[2] The new set of force constants was calculated for a cell determined with neutron diffraction at 295 K.[2] All mentioned lattice parameters are presented in Supplementary Table 7. Ionic positions of both structures were relaxed, prior to calculating the force constants, using the same MTP.

*Supplementary Table 7: Lattice parameters of the DFT-relaxed and experimentally determined crystal structure of naphthalene. The experiment was conducted at 295 K.*

| unit cell | $a$ / Å | $b$ / Å | $c$ / Å | $\beta$ / ° | $V$ / Å³ |
|---|---|---|---|---|---|
| DFT-relaxed[a] | 8.092 | 5.922 | 8.616 | 124.6 | 340.0 |
| exp. cell at 5 K[b] | 8.080 | 5.933 | 8.632 | 124.7 | 340.4 |
| exp. cell at 295 K[b] | 8.256 | 5.983 | 8.677 | 122.7 | 360.6 |

[a] dispersion-corrected DFT, ref.6
[b] Single crystal neutron diffraction at 5 K and 295 K, ref.2

The resulting (isotropic) thermal conductivities for both unit cell sizes are plotted in Supplementary Figure 21. They differ the most at low temperatures, e.g. by 0.42 Wm⁻¹K⁻¹ (41%) at 50 K, while towards room temperature this difference becomes far less significant. At 300 K the difference is only 0.03 Wm⁻¹K⁻¹ (8%).

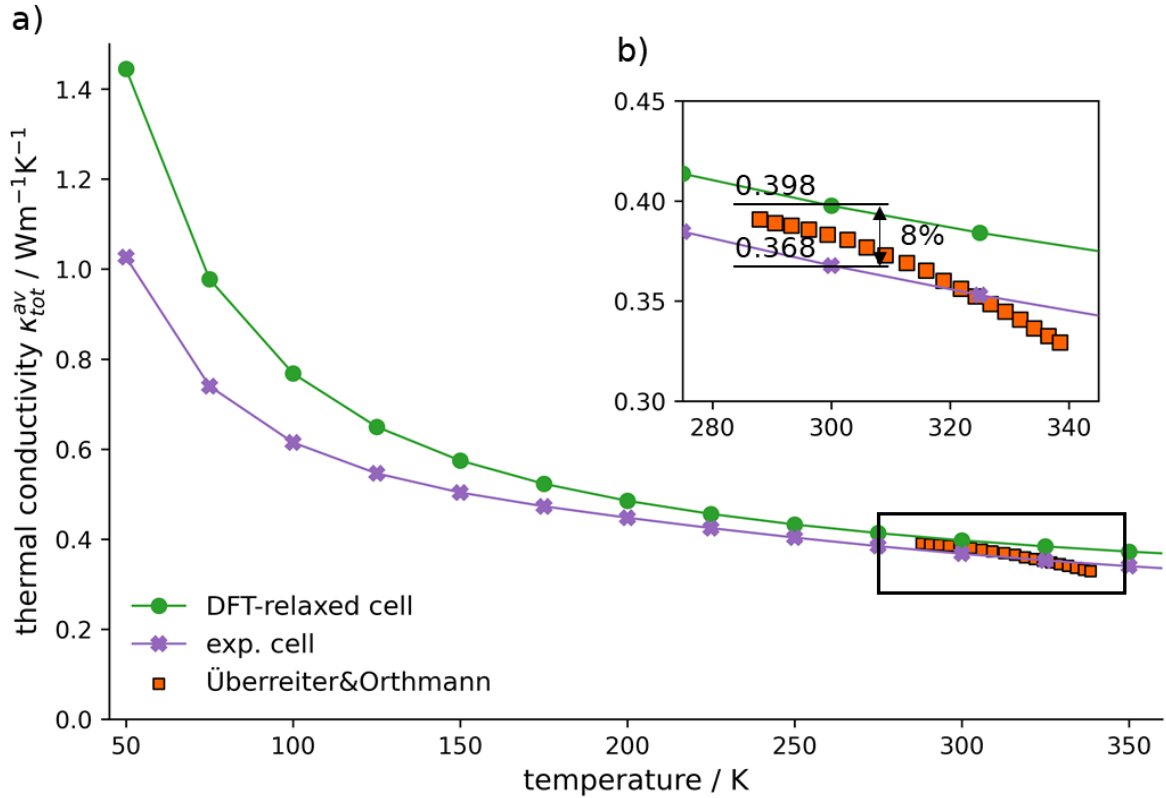

*Supplementary Figure 21: Isotropic, total lattice thermal conductivity of naphthalene as a function of temperature. a) The green line with dots shows the thermal conductivity of naphthalene with DFT-relaxed lattice parameters and the purple lines with crosses that of naphthalene with experimentally determined lattice parameters at 295 K. In panel b) a zoom of the region close to 300 K is presented. For comparison, the experimental data points of Überreiter and Orthmann[25] are printed as orange squares.*

The observation that the discrepancy is largest at low temperatures is not unexpected, as there, thermal conductivities are predominately determined by the acoustic phonons. Their frequencies are most strongly affected by intermolecular interactions, which in turn most strongly depend the intermolecular distances and, thus, on the unit cell sizes. Indeed, for the 295 K unit cell with the larger volume, the important phonon modes are shifted to lower frequencies (see Supplementary Figure 22), overall reducing the thermal conductivity.

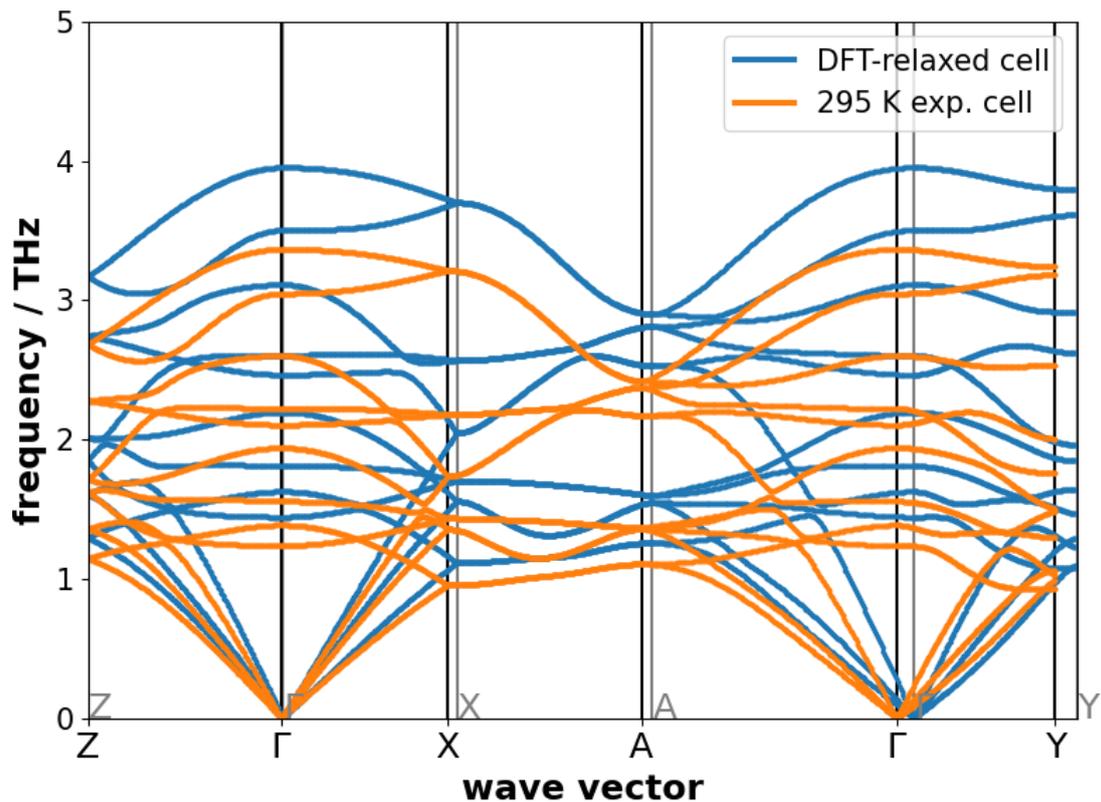

*Supplementary Figure 22: Phonon band structures that were calculated with the same system-specific MTP for naphthalene with either DFT-relaxed lattice parameters (blue bands) or experimentally determined lattice parameters at 295 K (orange bands). The difference of (Brillouin zone) volume is accounted for in the displayed lengths of the high-symmetry paths. The grey vertical lines denoting the high-symmetry paths and their grey labels correspond to the DFT-relaxed cell and their black counterparts to the 295 K cell.*

Our strategy to calculate the temperature-dependent thermal conductivities in the main work neglects thermal expansion as it uses the DFT-relaxed cell. However, this unit cell is highly suitable for low temperatures and at elevated temperatures, where its choice might appear no longer as ideal, only small deviations of $\kappa$ are observed in Supplementary Figure 21. Therefore, even though the approach to neglect thermal expansion in the calculations should be seen as an approximation, it doesn't change the trends for the relevant temperature region for the material at hand.

# Supplementary Section 5: Anisotropic lattice thermal conductivities of the acenes

The anisotropic thermal conductivities of the acenes for $T=300$ K are listed in Supplementary Table 8. For the entire considered temperature range between 50 K and 500 K, we refer to the provided output files in the repository.[1]

*Supplementary Table 8: Thermal conductivities of the four acenes at T=300 K. For each material the six independent thermal conductivity tensor elements (xx, yy, zz, yz, xz, xy) of the propagation ($\kappa_P$), the tunneling ($\kappa_C$), and the total ($\kappa_{tot}$) conductivity are presented individually.*

| acene | channel | xx / Wm⁻¹K⁻¹ | yy / Wm⁻¹K⁻¹ | zz / Wm⁻¹K⁻¹ | yz / Wm⁻¹K⁻¹ | xz / Wm⁻¹K⁻¹ | xy / Wm⁻¹K⁻¹ |
|---|---|---|---|---|---|---|---|
| | $\kappa_P$ | 0.235 | 0.404 | 0.346 | 0.000 | -0.018 | 0.000 |
| 2A | $\kappa_C$ | 0.065 | 0.095 | 0.110 | 0.000 | -0.018 | 0.000 |
| | $\kappa_{tot}$ | 0.300 | 0.499 | 0.456 | 0.000 | -0.036 | 0.000 |
| | $\kappa_P$ | 0.254 | 0.755 | 0.384 | 0.000 | -0.027 | 0.000 |
| 3A | $\kappa_C$ | 0.072 | 0.101 | 0.134 | 0.000 | -0.032 | 0.000 |
| | $\kappa_{tot}$ | 0.325 | 0.856 | 0.518 | 0.000 | -0.059 | 0.000 |
| | $\kappa_P$ | 0.241 | 0.383 | 0.315 | -0.039 | -0.051 | 0.006 |
| 4A | $\kappa_C$ | 0.074 | 0.113 | 0.215 | -0.021 | -0.039 | 0.011 |
| | $\kappa_{tot}$ | 0.315 | 0.497 | 0.530 | -0.060 | -0.089 | 0.017 |
| | $\kappa_P$ | 0.270 | 0.359 | 0.379 | 0.000 | -0.030 | -0.038 |
| 5A | $\kappa_C$ | 0.081 | 0.121 | 0.267 | -0.035 | -0.056 | 0.009 |
| | $\kappa_{tot}$ | 0.351 | 0.480 | 0.646 | -0.035 | -0.086 | -0.030 |

## Supplementary Section 5.1: Temperature-dependence of the anisotropic thermal conductivities

What follows are also further details on the anisotropy of the thermal conductivities in the acenes and its temperature-dependence. The results presented in this section are in analogy to the polar plots of directionally dependent thermal conductivities in the (001) plane, spanned by lattice vectors $a_1$ and $a_2$, and in the (010) plane, spanned by $a_1$ and $a_3$, in Figure 6d and Figure 6e in the main work. While those figures elucidate the directional-dependence at $T=300$ K, in Supplementary Figure 23 (panel a and b) analogous polar plots are shown for temperatures between 100 K and 500 K for pentacene. Here, the propagation conductivities $\kappa_P$ within the (001) plane in Supplementary Figure 23a and in the (010) plane in Supplementary Figure 23b are colored according to the corresponding temperature (see color bar). Additionally, the conductivities are superimposed on the crystal structure for the sake of clarity. Red stars (green crosses) indicate the maximum (minimum) for the respective temperature. For the (001) plane we see strong shifts in the positions of maximum and minimum of the thermal conductivities with temperature. To provide further context, the angle at which a maximum of $\kappa_{tot}$ is observed for each temperature is plotted as a function of temperature in Supplementary Figure 23c for the (001) plane and in Supplementary Figure 23d the (010) plane. These plots include the data for all acenes. The maximum within the (001) plane of pentacene shifts by almost 50°, while for tetracene it shifts by only 20°. Interestingly, the maxima of the two triclinic materials are not aligned, despite having almost the same angle $\gamma$ between $\boldsymbol{a_1}$ and $\boldsymbol{a_2}$ of 85.6° (4A) and 85.5° (5A). For the monoclinic acenes the tensor element $\kappa^{xy}(T)$ is zero, due to the symmetry and, therefore, the extrema in the (001) plane do not shift. However, in the (010) plane the maxima of each acene shift, due to $\kappa^{xz}(T)$ being always non-zero and due to its relative magnitude varying with $T$. Naphthalene, anthracene, and pentacene almost align towards higher temperatures, while for tetracene the shift is negligible. Notably, the maximum of neither acene is positioned at the corresponding lattice angles $\beta$, which are 124.6° (2A), 125.4° (3A), 113.2° (4A), and 112.7° (5A).

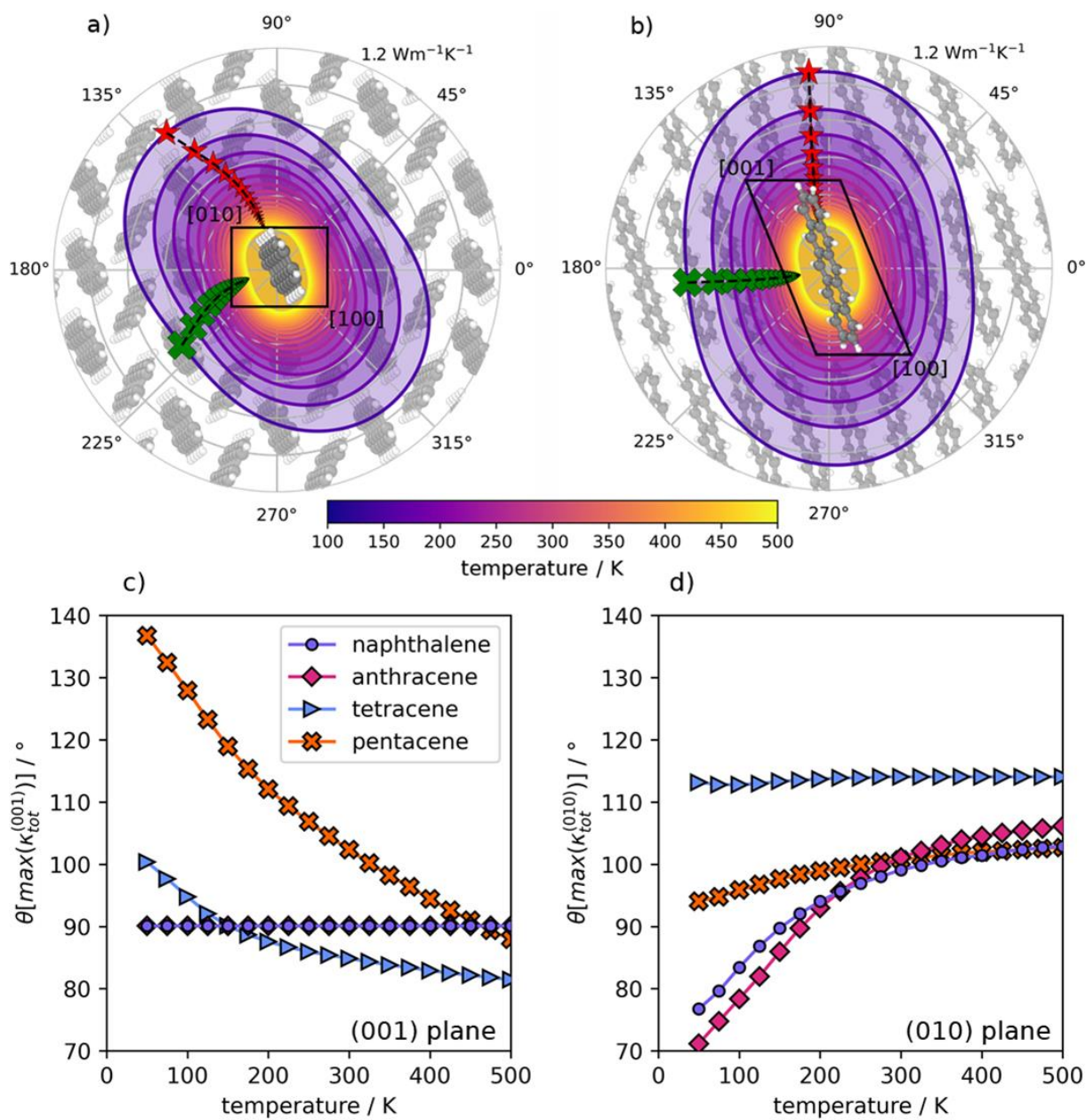

*Supplementary Figure 23: Polar plots of $\kappa_P(T)$ within the (001)-plane (spanned by lattice vectors $\boldsymbol{a_1}$ and $\boldsymbol{a_2}$) in a) and within the (010)-plane (spanned by $\boldsymbol{a_1}$ and $\boldsymbol{a_3}$) in b). The polar plots are superimposed over the crystal structures of pentacene and the respective crystal directions are labeled. The gray rings indicate increments of 0.2 $Wm^{-1}K^{-1}$. The color bar indicates the temperature. The red stars (green crosses) mark the position of the maximum (minimum) at each temperature. In panels c) and d) the positions of those maxima are plotted (linearly) as a function of temperature for each acene.*

## Supplementary Section 5.2: Impact of anharmonicity on the anisotropic thermal conductivities

As a "rough" test to assess to what extent increasing the degree of anharmonicity would impact the observed trends, we calculated the thermal conductivities of the smallest (naphthalene) and the largest acene (pentacene) with rescaled phonon linewidths to test how much the conductivities are influenced by the anharmonicity.[27] In Supplementary Figure 24 the linewidths $\Gamma_{anh}$ are either decreased by a factor of 1.25 (dashed line), unchanged (solid), or increased by 1.25 (dotted) and $\kappa_{tot}$ and $\kappa_C$ are individually shown for naphthalene (blue) and pentacene (orange). Overall, one sees that none of the discussed trends are affected by the rescaling and therefore, the effective phonon linewidth. Only in 5A in the [001] direction the temperature at which the minimum thermal conductivity occurs is shifted. This is a consequence of $\kappa_C$ exhibiting a higher dependence on the phonon linewidth in [001] than in the other directions.

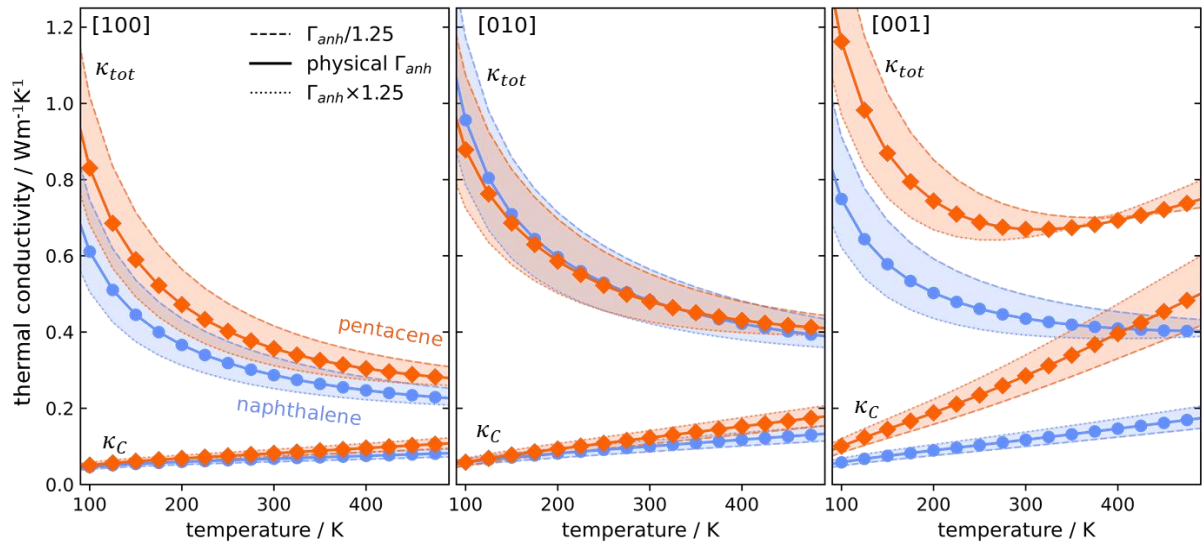

*Supplementary Figure 24: Anisotropic lattice thermal conductivities of naphthalene (blue) and pentacene (orange) as a function of temperature. Solid lines show the total thermal conductivity $\kappa_{tot}$ and $\kappa_C$ according to their physical anharmonic linewidths $\Gamma_{anh}$, while dashed and dotted lines show conductivities where the anharmonicity was decreased or increased by a factor of 1.25. The anisotropic conductivities are presented for the crystal directions [100] in panel a), for [010] in panel b), and for [001] in panel c), according to eq.(3) in the main work.*

## Supplementary Section 6: Acoustic phonons in the acene crystals

In the main text, we refer to the acoustic phonons bands several times and although, they are in principle a well-understood concept, we want to provide an illustration of those phonon bands for the acenes. For that purpose, we use the "acoustic participation ratio" (APR)[28] to color the phonon bands in Supplementary Figure 25. The definition of the APR can be found in the Supplementary Material of ref.26. It is a measure for the spatial alignment of the atomic eigendisplacements of a phonon mode. Therefore, the APR of a phonon mode $\lambda$ is defined in equation S1 by the sum of the scalar products of atomic eigenvectors $e_\lambda$ between atom $i$ and $j$ within the unit cell and their respective mass $m$. Here, mode $\lambda$ refers to a specific band at a specific wave vector.

$$APR_\lambda = \frac{2}{N(N+1)} \frac{\left| \sum_{i=1}^{3N} \sum_{j \geq i}^{3N} \frac{\left(e_\lambda^i\right)^\dagger e_\lambda^j}{\sqrt{m_i m_j}} \right|^2}{\sum_{i=1}^{3N} \sum_{j \geq i}^{3N} \left| \frac{\left(e_\lambda^i\right)^\dagger e_\lambda^j}{\sqrt{m_i m_j}} \right|^2} \qquad (S1)$$

If all atoms are displaced in the same direction, the APR equals 1 because it is weighted by a factor that considers the number of atoms per unit cell $N$. In Supplementary Figure 25, we apply the APR to identify the phonon bands with acoustic character. For the acenes, acoustic character means that both molecules move in phase as rigid bodies. All other vibrations form the optical bands and those either represent anti-phase rigid-body translations or they are localized on individual atoms or moieties. In Figure 25 colors from dark red to light blue show that a band has at least some degree of acoustic character, while dark blue is reserved for the optical modes (see color bar above the panels). In this way, it is possible to track the acoustic phonons from the $\Gamma$-point towards the Brillouin zone boundaries. The data also show that the mostly linear dispersion of an acoustic band persists, despite avoided crossings with optical bands. At such avoided crossings, the acoustic character simply passes on to the next band, which is reflected by the color within a band abruptly changing between dark red and blue.

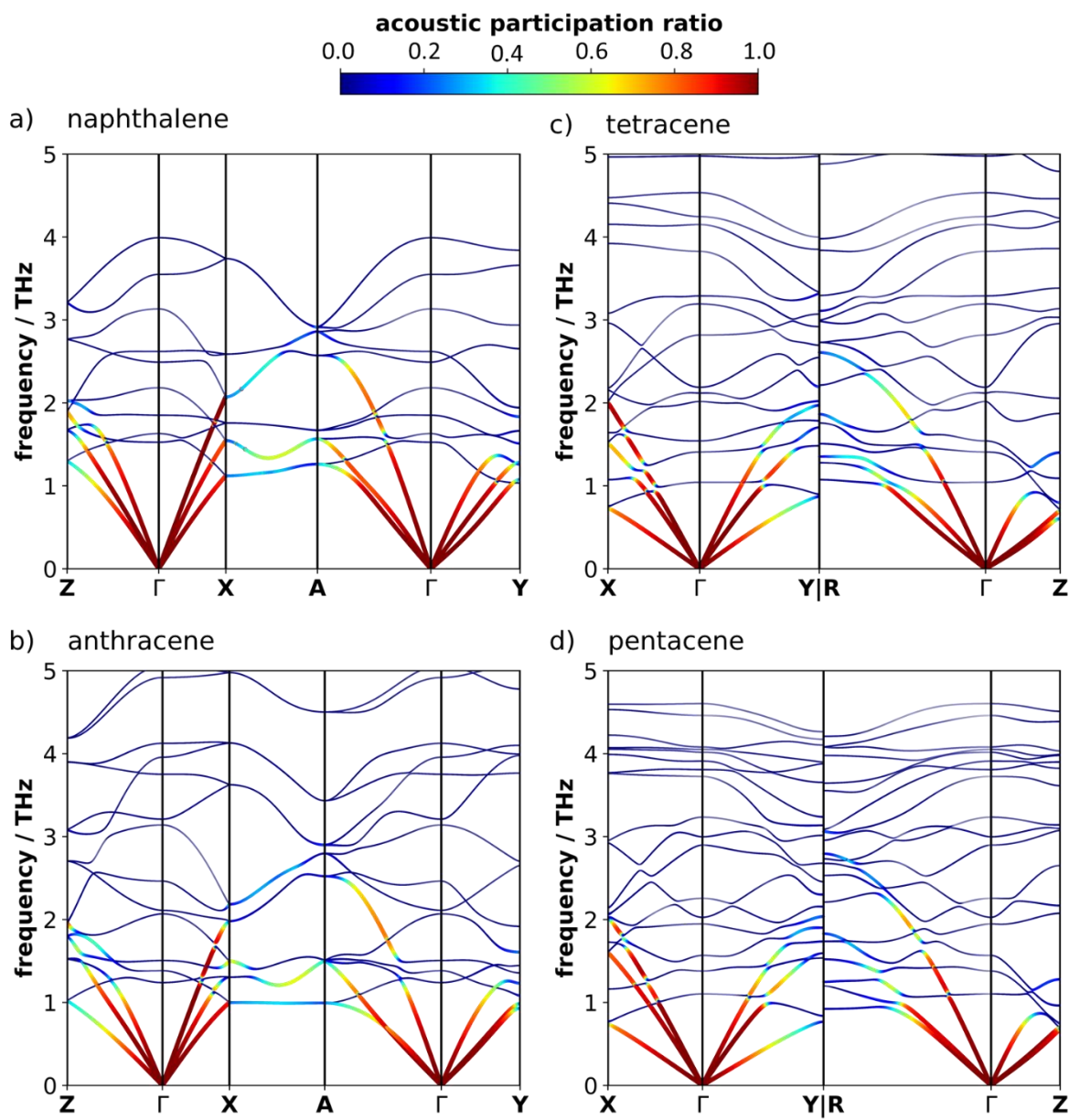

*Supplementary Figure 25: Low-frequency phonon band structures of naphthalene a), anthracene b), tetracene c), and pentacene d) calculated with their system-specific MTPs and colored according to their normalized acoustic participation ratio (APR). (Primarily) acoustic modes are characterized by colors from dark red to light blue (see color bar above the panels), while the modes colored in dark blue form optical bands.*

# Supplementary References